\documentclass[10pt,twocolumn,journal]{IEEEtran}
\usepackage{latexsym}
\usepackage{float}
\usepackage{amsfonts}
\usepackage{amsbsy}
\usepackage{amssymb}
\usepackage{times}
\usepackage{graphicx}
\usepackage{enumerate}
\usepackage[usenames]{color}
\usepackage[dvips]{pstcol}
\usepackage{epstopdf}
\usepackage{cite}
\usepackage{amssymb}
\usepackage{amsfonts}
\usepackage{graphicx}
\usepackage{epsfig}
\usepackage{psfrag}
\usepackage{xcolor}
\usepackage{amsfonts, bm}
\usepackage{epstopdf}
\usepackage{cite}
\usepackage{color}
\usepackage{xcolor}
\usepackage{subfig}
\usepackage{verbatim}
\usepackage{multirow}
\usepackage{array}
\usepackage{booktabs}
\usepackage{amsthm}

\usepackage{algorithm}
\usepackage{algorithmicx}
\usepackage{algpseudocode}
\usepackage{amsmath}

\newtheorem{theorem}{Theorem}
\newtheorem{lemma}{Lemma}

\IEEEoverridecommandlockouts
\begin{document}
\title{ {\huge DDPG-Driven Deep-Unfolding with Adaptive Depth for Channel Estimation with Sparse Bayesian Learning } }
\author{ Qiyu Hu, \textit{Student Member, IEEE,} Shuhan Shi, Yunlong Cai, \textit{Senior Member, IEEE,} \\ and Guanding Yu, \textit{Senior Member, IEEE}  
	\thanks{
		Q. Hu, S. Shi, Y. Cai, and G. Yu are with the College of Information Science and Electronic Engineering, Zhejiang University, Hangzhou 310027, China (e-mail: qiyhu@zju.edu.cn; ssh16@zju.edu.cn; ylcai@zju.edu.cn; yuguanding@zju.edu.cn).
	}
}

\maketitle
\vspace{-3.3em}
\begin{abstract}
Deep-unfolding neural networks (NNs) have received great attention since they achieve satisfactory performance with relatively low complexity. Typically, these deep-unfolding NNs are restricted to a fixed-depth for all inputs. 
However, the optimal number of layers required for convergence changes with different inputs. 
In this paper, we first develop a framework of deep deterministic policy gradient (DDPG)-driven deep-unfolding with adaptive depth for different inputs, where the trainable parameters of deep-unfolding NN are learned by DDPG, rather than updated by the stochastic gradient descent algorithm directly. Specifically, the optimization variables, trainable parameters, and architecture of deep-unfolding NN are designed as the state, action, and state transition of DDPG, respectively. 
Then, this framework is employed to deal with the channel estimation problem in massive multiple-input multiple-output systems. Specifically, first of all we formulate the channel estimation problem with an off-grid basis and develop a sparse Bayesian learning (SBL)-based algorithm to solve it. Secondly, the SBL-based algorithm is unfolded into a layer-wise structure with a set of introduced trainable parameters. Thirdly, the proposed DDPG-driven deep-unfolding framework is employed to solve this channel estimation problem based on the unfolded structure of the SBL-based algorithm. To realize adaptive depth, we design the halting score to indicate when to stop, which is a function of the channel reconstruction error.
Furthermore, the proposed framework is extended to realize the adaptive depth of the general deep neural networks (DNNs). Simulation results show that the proposed algorithm outperforms the conventional optimization algorithms and DNNs with fixed depth with much reduced number of layers.
\end{abstract}

\IEEEpeerreviewmaketitle

\section{Introduction}
\subsection{Prior Work}
Recently, the deep learning have attracted great attention and have been widely employed in wireless communications due to their satisfactory performance and low complexity \cite{DLMaga}. 
The deep neural networks (DNNs) have been applied to symbol detection \cite{Viterbi}, beamforming \cite{LearnOpt}, channel feedback \cite{CSIFeed,CSIFeed2}, and channel estimation \cite{DNNchannel1,DNNchannel2,DNNchannel3}. In particular, channel correlation are captured by DNNs to improve the accuracy of channel estimation \cite{DNNchannel1,DNNchannel2,DNNchannel3}. 
In general, the black-box DNNs have poor interpretability and generalization ability, and require a sufficiently large number of training samples. To address these issues, deep-unfolding neural networks (NNs) are proposed in \cite{UnfoldSurvey,UnfoldWMMSE,AMP,CSIEstUnfold}, where the iterative algorithms are unfolded into a layer-wise architecture similar to DNNs \cite{UnfoldSurvey}. Specifically, the classical weighted minimum mean-square error algorithm has been unfolded in \cite{UnfoldWMMSE} for beamforming. An approximate message passing based deep-unfolding NN has been proposed in \cite{AMP}. In \cite{CSIEstUnfold}, an alternative direction method of multipliers algorithm has been unfolded for channel estimation. 
Moreover, to solve the mixed integer non-linear problems, deep Q-network (DQN) \cite{Nature} has been widely employed in communications, such as beam selection \cite{BeamSelect}, dynamic channel access \cite{DRLAccess}, and resource allocation \cite{DRLresou}, where the problems are modeled as Markov decision process (MDP). In addition, a sort of DQN referred to as deep deterministic policy gradient (DDPG) has been developed to solve the problems with continuous actions \cite{DDPG0,TuningFree,DDPG1,DDPG2}. It has been employed for power control in device-to-device communications \cite{DDPG1} and wireless power transfer \cite{DDPG2}.
 
\begin{figure}[t]
\begin{centering}
\includegraphics[width=0.5\textwidth]{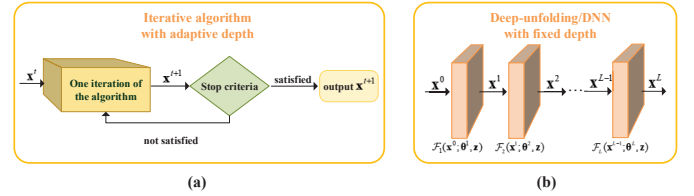}
\par\end{centering}
\caption{(a) Iterative algorithms with adaptive depth; (b) Deep learning-based algorithms with fixed depth.}
\label{LearnToStop}
\end{figure}

However, most of the studies on deep-unfolding focus on improving performance, and neglects a main difference between the deep-unfolding NN and iterative algorithms \cite{AdaptDepth}, where the latter can adjust the number of iterations for different inputs, as presented in Fig. \ref{LearnToStop}(a). In contrast, the depth of deep-unfolding NN is pre-determined and the computational complexity is proportional to the depth, as shown in Fig. \ref{LearnToStop}(b) \cite{ACT}. The fixed depth generally results in two defects \cite{Timedata}: (i) The waste of computing resources when DNN is employed to handle ``easy" samples; (ii) The unsatisfied performance when DNN is used for ``hard" samples.
Moreover, pursuing better performance is not the only target and some applications require to solve the problem within a given time. The 5G communication systems have strict requirements for latency \cite{ITUR}, hence algorithms with high complexity are impractical. To this end, a promising field of growing interests is to propose fast yet accurate deep-unfolding NNs with adaptive depth \cite{ACT,TuningFree,AdaptDepth,Timedata}.
However, there are some open issues to be addressed in the existing deep-unfolding structure with adaptive depth: (i) They generally have a discontinuous loss function, which makes it difficult to be trained \cite{ACT}; (ii) The halting probability of each layer is computed only based on the output of that layer, which results in a bad choice of the number of layers \cite{TuningFree}; (iii) The output of DNN is a weighted sum of outputs from all layers, which is different from iterative algorithms that output the results from last iteration \cite{ACT}; (iv) The maximum number of layers of DNN need to be pre-determined and the DNN is trained with the maximum number of layers, which leads to severe gradient vanishing or explosion \cite{AdaptDepth}.

In wireless communications, massive multiple-input multiple-output (MIMO) has attracted great attention and has been considered as a key technology to meet the capacity requirements in beyond 5G wireless systems \cite{MIMO1,MIMO2,MIMO3}. To take full advantage of sufficient base station (BS) antennas, it is important to know channel state information (CSI) at the transmitter \cite{CSIAcq}. Fortunately, since the number of scatterers is limited in the propagation environment, the massive MIMO channel has an approximately sparse representation under the discrete Fourier transform (DFT) basis \cite{DFT1,DFT2,OMP}. By exploiting such sparsity, there are lots of efficient channel estimation methods \cite{TwoStage,SparsityAlgo1,SparsityAlgo2}. 
However, the DFT-based channel estimation algorithms generally have a performance loss due to the leakage of energy in the DFT basis \cite{Dictionary1}. Actually, the DFT basis provides fixed sampling grids which discretely cover the angular domain of channel. Since signals generally come from random directions, leakage energy caused by direction mismatch is inevitable. To reduce the leakage energy, the authors in \cite{Dictionary1} considered an over-complete DFT basis, which provides denser sampling grids on the angular domain. The over-complete DFT basis still results in severe direction mismatch, when the grid is not dense enough. In addition, if a very dense sampling grid is employed, the performance of recovery algorithms degrades since the basis vectors are highly correlated. 
To overcome these drawbacks, the authors in \cite{SBLChannel} introduced an off-grid model for channel sparse representation and a sparse Bayesian learning (SBL) approach has been proposed in \cite{SBLSense} for channel recovery. Though the existing algorithms achieve satisfactory performance, they generally have high computational complexity and some a priori parameters are hard to determine, which poses great challenges in practical implementations.

\subsection{Motivation and Contribution}
To the best of our knowledge, deep-unfolding with adaptive depth has not been well investigated in multiuser MIMO (MU-MIMO) systems. In this paper, we firstly develop a framework of DDPG-driven deep-unfolding NN with adaptive depth. The trainable parameters of deep-unfolding NN are learned by DDPG, rather than updated by stochastic gradient descent (SGD) algorithm directly, which avoids the gradient vanishing and explosion. 
In particular, the increase of deep-unfolding depth has severe impacts on the back propagation process of gradients, which leads to the performance degradation. It is difficult to be well solved due to the complex architecture and non-linear function of deep-unfolding. This problem can be solved well by DDPG unfolding, since the increase of deep-unfolding depth simply leads to the increase of time steps in DDPG and it will not effect its architecture and the back propagation process of gradients.
Specifically, the optimization variables, trainable parameters, and architecture of deep-unfolding NN are designed as the state, action, and state transition of DDPG, respectively. To achieve adaptive depth, we carefully design the reward function and halting score. 
The motivations to employ DDPG mainly lie in: (i) The DDPG can achieve adaptive depth for deep-unfolding; (ii) It can avoid gradient vanishing and explosion; (iii) The DDPG can deal with the continuous actions \cite{DDPG0}, where the actions. i.e., trainable parameters, are generally continuous. 

This framework is employed to solve the channel estimation problem in massive MIMO systems. A general off-grid model is considered for channel sparse representation of massive MIMO systems with uniform linear array (ULA). Then, we formulate the problem of channel estimation with an off-grid basis and employ a block majorization-minimization (MM) approach \cite{MMAlgorithm} for joint sparse channel recovery and off-grid refinement. In particular, an efficient SBL-based \cite{SBLChannel} algorithm is proposed to refine the grid points, off-grid variables, and parameters in prior distribution iteratively. 
Subsequently, the SBL-based algorithm is unfolded into a layer-wise structure to achieve the performance approaching the SBL-based algorithm with a much smaller number of layers. To improve the accuracy of channel estimation, a set of trainable parameters are introduced, which can be divided into two categories: (i) Existing parameters in the SBL-based algorithm, such as a priori parameters that are difficult to determine; (ii) The introduced trainable parameters to replace the operations with high computational complexity.

Different channel samples have various sparsity level, hence generally require different numbers of iterations of SBL-based algorithm (deep-unfolding layers). Thus, according to the unfolded structure of the SBL-based algorithm, we employ the proposed DDPG-driven deep-unfolding framework with adaptive depth to solve this channel estimation problem. In particular, the optimization variables, i.e., grid points, off-grid variables, and parameters in prior distribution, are defined as state. The introduced trainable parameters are treated as action and the deep-unfolding architecture of the SBL-based algorithm is regarded as the state transition. 
To realize the adaptive depth, we design the halting score to indicate when to stop, which is a function of the channel reconstruction error. In particular, a DNN is designed to learn the halting score, which can be treated as a sub-network of DDPG. The reward function of DDPG is designed as the weighted sum of two parts: (i) The decrease of normalized mean square error (NMSE), i.e., performance improvement between two iterations, and a penalty is introduced in the reward to penalize the policy if the DDPG does not select to terminate. Hence, a negative reward will be given if the performance improvement cannot exceed the penalty, thus forcing the policy to early stop with diminished reward. It ensures that the channel reconstruction error significantly decreases in each layer. (ii) The function related to the halting score, where we can control the channel reconstruction error and the number of layers by tuning the introduced hyper-parameters.

The main contributions of this paper are summarized as follow:
\begin{itemize}
\item We develop a framework of DDPG-driven deep-unfolding with adaptive depth for different inputs, where the trainable parameters are learned by the DDPG.

\item We formulate the channel estimation problem in massive MIMO systems with an off-grid basis and develop the SBL-based algorithm to solve it. The SBL-based algorithm is unfolded into a layer-wise structure and a set of trainable parameters are introduced to improve the performance. The performance analysis is provided for this deep-unfolding NN.

\item The proposed DDPG-driven deep-unfolding framework is employed to solve this problem based on the unfolded structure of the SBL-based algorithm. We design a halting score to realize the adaptive depth, which is a function of the channel reconstruction error. 

\item The proposed framework is extended to realize the adaptive depth of the general DNNs. Simulation results show that our proposed DDPG-driven deep-unfolding significantly outperforms conventional optimization algorithms and DNNs with fixed depth in terms of the NMSE performance with much reduced number of layers.
\end{itemize}

\subsection{Organization and Notations}
The rest of the paper is structured as follows. Section \ref{Framework} proposes a general framework of DDPG-driven deep-unfolding with adaptive depth. Section \ref{SBLAlgorithm} formulates the sparse channel estimation problem and develops an efficient SBL-based algorithm. Section \ref{DeepUnfold} unfolds the SBL-based algorithm into a layer-wise structure and provides the performance analysis. Section \ref{DDPGChannel} proposes the DDPG-driven deep-unfolding with adaptive depth based on the unfolded SBL-based algorithm.
The simulation results are presented in Section \ref{Simulation}. Finally, the paper is concluded in Section \ref{Conclusion}.

\emph{Notations:} Scalars, vectors, and matrices are respectively denoted by lower case, boldface lower case, and boldface upper case letters.
The notation $\mathbf{I}$ represents an identity matrix and $\mathbf{0}$ denotes an all-zero matrix.
For a matrix $\mathbf{A}$, ${\bf{A}}^T$, $\mathbf{A}^*$, ${\bf{A}}^H$, ${\bf{A}}^{-1}$, ${\bf{A}}^{\dagger}$, and $\|\mathbf{A}\|$ are its transpose, conjugate, conjugate transpose, inversion, pseudo-inversion, and Frobenius norm, respectively.
For a vector $\mathbf{a}$, $\|\mathbf{a}\|$ is its Euclidean norm.
We use $\mathbb{E}\{ \cdot \}$ for the statistical expectation, $\Re\{ \cdot \}$ ($\Im\{ \cdot \}$) denotes the real (imaginary) part of a variable, $\textrm{Tr}\{ \cdot \}$ is the trace operation, $| \cdot |$ denotes the absolute value of a complex scalar, and $\circ$ is the element-wise multiplication of two matrices, i.e., Hadmard product.
Finally, ${\mathbb{C}^{m \times n}}\;({\mathbb{R}^{m \times n}})$ are the space of ${m \times n}$ complex (real) matrices.

\section{The Proposed DDPG-Driven Deep-Unfolding Framework} \label{Framework}
In this section, we propose the framework of DDPG-driven deep-unfolding with adaptive depth.

\subsection{Problem Setup}

\subsubsection{Optimization Problem and Iterative Algorithm}
An optimization problem has the following general form
\begin{equation}
\min\limits_{\mathbf{x}} \quad f(\mathbf{x};\mathbf{z}) \quad \text{s.t.} \quad \mathbf{x}\in \mathcal{X}, \label{generalform}
\end{equation}
where $f: \mathbb{C}^{m}\mapsto \mathbb{R}$ denotes the objective function, $\mathbf{x}\in \mathbb{C}^{m}$ is the optimization variable, $\mathcal{X}$ denotes the feasible region, and $\mathbf{z}\in \mathbb{C}^{p}$ is the pre-determined parameter.

As presented in Fig. \ref{LearnToStop}(a), an iterative algorithm is proposed to solve problem \eqref{generalform} as
\begin{equation}
\mathbf{x}^{t}=F_{t}(\mathbf{x}^{t-1};\mathbf{z}),  \label{generalsolu}
\end{equation}
where $t\in \mathcal{T}\triangleq \{1, 2, \ldots , T_{a}\}$ is the index of iteration, $T_{a}$ denotes the maximum number of iterations, and function $F_{t}$ maps variable $\mathbf{x}^{t-1}$ to $\mathbf{x}^{t}$ at the $t$-th iteration based on the parameter $\mathbf{z}$.

\subsubsection{Deep-Unfolding}
Deep-unfolding NN is developed to unfold the iterative algorithm into a layer-wise structure. Based on the iteration expression \eqref{generalsolu}, by introducing the trainable parameter $\bm{\theta}\in \mathbb{C}^{a\times b}$, a deep-unfolding NN is proposed in Fig. \ref{LearnToStop}(b) as
\begin{equation}
\mathbf{x}^{l}=\mathcal{F}_{l}(\mathbf{x}^{l-1};\bm{\theta}^{l},\mathbf{z}), \label{generalnetwork}
\end{equation}
where $l\in \mathcal{L}\triangleq \{1, 2, \ldots , L\}$ denotes the index of layer in the deep-unfolding NN, $L$ is the total number of layers, $\mathcal{F}_{l}$ denotes the structure of deep-unfolding NN in the $l$-th layer, $\mathbf{x}^{l-1}$ and $\mathbf{x}^{l}$ represent the input and output of the $l$-th layer, respectively, $\mathbf{z}$ is the given parameter, i.e., input of the deep-unfolding NN, and $\bm{\theta}^{l}$ denotes the introduced trainable parameter in the $l$-th layer.
Furthermore, the objective function $f(\mathbf{x};\mathbf{z})$ in \eqref{generalform} can be treated as the loss function.

\subsection{Introduction of DDPG}

\begin{figure}[t]
\begin{centering}
\includegraphics[width=0.45\textwidth]{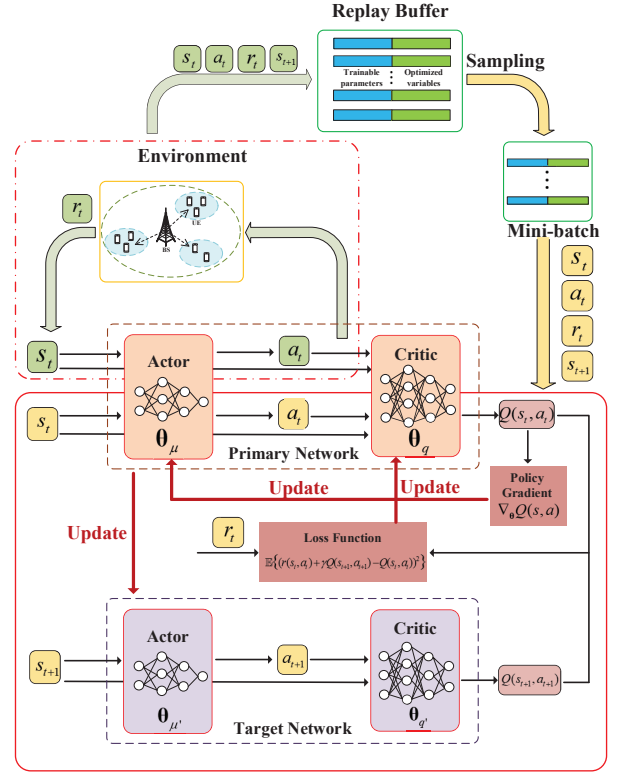}
\par\end{centering}
\caption{DDPG architecture.}
\label{DDPGFramework}
\end{figure}

\subsubsection{MDP}

\begin{figure*}[t]
\begin{centering}
\includegraphics[width=0.8\textwidth]{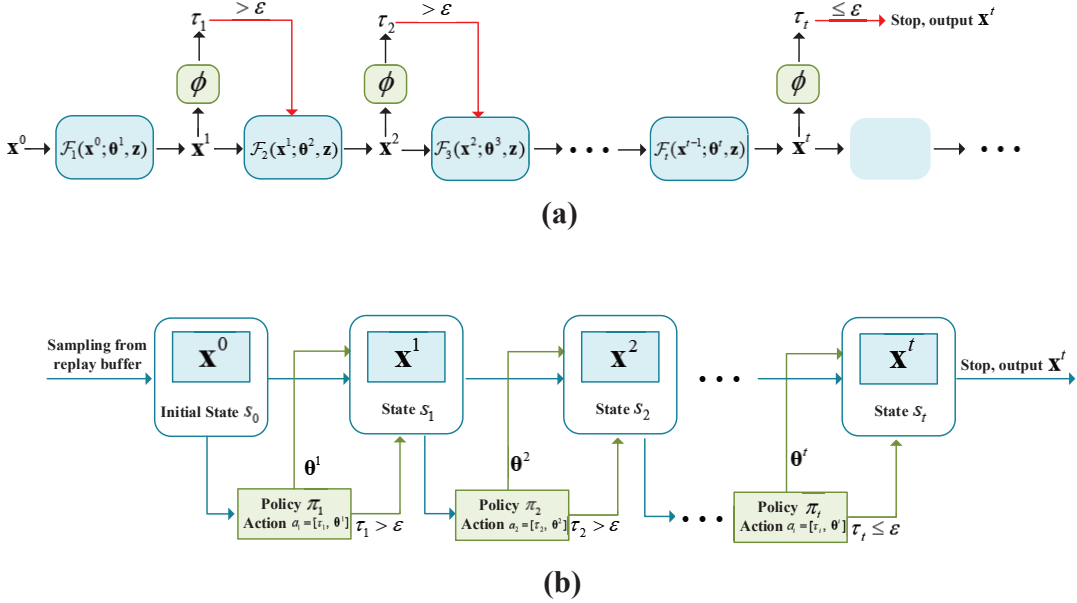}
\par\end{centering}
\caption{(a) Deep learning-based algorithms with adaptive depth; (b) Framework of DDPG-driven deep-unfolding with adaptive depth.}
\label{StopPredict}
\end{figure*}

The MDP is defined by the quintuple $(\mathcal{S}, \mathcal{A}, \mathcal{R}, \mathcal{P}, \gamma)$, where $\mathcal{S}$ represents the state space, $\mathcal{A}$ is the action space, $\mathcal{R}: \mathcal{S}\times \mathcal{A} \rightarrow \mathbb{R}$ is the reward function, $\mathcal{P}: \mathcal{S} \times \mathcal{A} \rightarrow \mathcal{S}$ denotes the state transition function, and $\gamma \in (0,1)$ is the discount factor. 
Correspondingly, $s_{t}$, $a_{t}$, and $r_{t}$ denote state, action, and reward at time step $t$, respectively.
We denote $p$ as the transition probability and $\pi$ as the policy, $\pi: \mathcal{S} \mapsto \mathcal{A}$. Then, we have $a_{t}\sim \pi(\cdot|s_t)$, $s_{t+1}\sim p(\cdot|s_{t}, a_{t})$, $r_{t}\triangleq r(s_{t}, a_{t})\sim \mathcal{R}$, and the cumulative discounted reward $\mathbb{E} \{ \sum_{t\geq 0}\gamma^{t}r_{t} \}$. 

\subsubsection{DDPG} 
DDPG is a hybrid model that combines the methods of value function and policy search \cite{BeamSelect,DRLAccess,DRLresou,DDPG0,TuningFree,DDPG1,DDPG2}. 
Taking advantages of both models, DDPG improves the convergence speed to be suitable for large-scale action spaces and can deal with continuous actions.
As presented in Fig. \ref{DDPGFramework}, DDPG consists of two basic elements \cite{DDPG0}: actor and critic. Specifically, the actor function $\pi_{\bm{\theta}_{\mu}}(a|s)$ maps the state into a specific action based on the current policy and the critic function $Q(s, a)$ is learned by Q-learning to evaluate the selected action. We employ experience replay buffer $\mathcal{D}$ \cite{DDPG1} and DDQN architecture \cite{DDPG2} for the actor and critic, respectively. 
In particular, a mini-batch of transitions $(s_{t}, a_{t}, r_{t}, s_{t+1})$ are selected from $\mathcal{D}$ by random sampling for training. The trainable parameters of critic network $\bm{\theta}_{q}$ are updated by SGD and the actor policy is updated by employing the sampled policy gradient \cite{DDPG0}. 
The parameters of the target actor NN $\bm{\theta}_{\mu'}$ and the target critic NN $\bm{\theta}_{q'}$ are copied from those of the main actor NN $\bm{\theta}_{\mu}$ and the main critic NN $\bm{\theta}_{q}$ once in a while, respectively.

\subsection{DDPG-Driven Deep-Unfolding with Adaptive Depth} \label{FrameDDPG}

Fig. \ref{StopPredict}(a) shows the idea of deep-unfolding NN with adaptive layers \cite{AdaptDepth}, where $\phi$ denotes the function that outputs $\tau_{t}$ to indicate whether to halt at the $t$-th layer. Based on this, we further propose the DDPG-driven deep-unfolding with adaptive depth. In particular, a layer of the deep-unfolding is modeled as a transition of the MDP, as shown in Fig. \ref{StopPredict}(b). The training of deep-unfolding is to optimize trainable parameters $\bm{\theta}^{l}, \forall l$ in each layer, which can be treated as the action of DDPG. We formulate the MDP as below. To avoid confusion, we use $l$ and $t$ to denote the index of layers in deep-unfolding NN and time steps in DDPG, respectively.
\begin{itemize}
\item Agent: The BS observes the state $s_t$ and selects an action $a_t$ according to the policy $\pi$ to interact with the environment. Then, the environment feeds back the reward and the BS adjusts its policy $\pi$ correspondingly. We aim to learn the optimal policy $\pi$ to maximize the cumulative discounted reward $\mathbb{E}\big\{ \sum_{t\geq 0}\gamma^{t} r_{t} \big\}$.

\item State space: $\mathcal{S}$ denotes the space of optimization variables, which consists of the initial value  $\mathbf{x}^{0}$ and all intermedia results $\mathbf{x}^{l}, \forall l$ in the optimization process, i.e., the output of each layer of deep-unfolding NN. The state at the $t$-th time step $s_{t}$ is composed of the output of deep-unfolding NN in the $t$-th layer, i.e., $s_{t}\triangleq \mathbf{x}^{t}$.

\item Action space: $\mathcal{A}$ is composed of the halting indicator $\tau$ and the trainable parameters $\bm{\theta}^{l}, \forall l$ in each layer. The action at the $t$-th time step consists of $\tau_{t}$ and the trainable parameters of the $t$-th layer in the deep-unfolding NN, i.e., $a_{t}\triangleq [\tau_{t}, \bm{\theta}^{t}]$. The role of $\tau_{t}\in [0,1]$ is to determine whether to halt the running of deep-unfolding NN at the current layer. We move forward to the next iteration (layer) if $\tau_{t} > \varepsilon$, where $\varepsilon$ is a hyper-parameter. Otherwise the running of deep-unfolding NN would be halted to output the final state as results. 

\item State transition: The transition function $\mathcal{P}: \mathcal{S} \times \mathcal{A} \rightarrow \mathcal{S}$ maps the current state $s_{t}$ to the next state $s_{t+1}$ based on the selected action $a_{t}$. The state transition $s_{t+1}=p(s_{t}, a_{t})$ is composed of one or several layers of the deep-unfolding NN, i.e., $\mathbf{x}^{l+1} = \mathcal{F}_{l+1}(\mathbf{x}^{l};\bm{\theta}^{l+1},\mathbf{z})$ in \eqref{generalnetwork}. 

\item Reward: After each transition, the environment feeds back a reward $r_{t}$ according to the reward function $\mathcal{R}: \mathcal{S}\times \mathcal{A} \rightarrow \mathbb{R}$, which is designed as the performance improvement between the former and current layers, i.e., $r_{t}=f(\mathbf{x}^{t-1};\mathbf{z})-f(\mathbf{x}^{t};\mathbf{z})-\eta$. Note that $f(\mathbf{x}^{t};\mathbf{z})$ denotes the objective function in \eqref{generalform} at the $t$-th iteration and a higher reward is received when the policy results in higher performance improvement. Moreover, $\eta$ is a constant and it penalizes the policy as it does not select to halt at time step $t$. A negative reward will be given if the performance improvement cannot exceed the penalty $\eta$, thus forcing the policy to early stop with diminished reward.
\end{itemize}

\section{SBL-Based Algorithm for Channel Estimation} \label{SBLAlgorithm}

In this section, we introduce a model-based off-grid basis to deal with the direction mismatch. Then, this off-grid model is applied to the channel estimation and we formulate the problem accordingly. We model the distributions of the off-grid parameters and develop the SBL-based algorithm to solve this problem. 

\subsection{An Off-Grid Basis for Massive MIMO Channels}
We consider a flat fading channel. The downlink channel vector from the BS to the $k$-th user that consists of $N_{c}$ clusters with $N_{s}$ propagating rays is given by \cite{SBLChannel} 
\begin{equation} \label{ChannelModel}
\mathbf{h}_{k}=\sum\limits_{i=1}^{N_{c}}\sum\limits_{j=1}^{N_{s}}\xi_{ij}\mathbf{a}(\phi_{ij}),
\end{equation}
where $\xi_{ij}\sim \mathcal{CN}(0,\sigma_{\alpha}^{2})$ is the complex gain of the $j$-th ray in the $i$-th cluster, $\phi_{ij}$ denotes the corresponding angles-of-departure (AoD), and $\mathbf{a}(\phi_{ij})$ represents the array response vectors. For a ULA with $N$ antennas, the response vector is given by
\begin{equation}
\mathbf{a}(\phi) = \frac{1}{\sqrt{N}}\big[1, e^{-j 2\pi \frac{d}{\lambda} \sin(\phi)}, \cdots , e^{-j 2\pi \frac{d}{\lambda}(N-1) \sin(\phi)}  \big]^{T},
\end{equation}
where $d$ and $\lambda$ denote the distance between the adjacent antennas and carrier wavelength, respectively.

The BS is equipped with a ULA and it transmits a sequence of $T$ pilot symbols, denoted by $\mathbf{X}\in \mathbb{C}^{T\times N}$, for each user to perform channel estimation. Thus, the received signal $\mathbf{y}_{k}\in \mathbb{C}^{T\times 1}$ at the $k$-th user is
\begin{equation} \label{SignalModel}
\mathbf{y}_{k} = \mathbf{X}\mathbf{h}_{k} + \mathbf{n}_{k},
\end{equation}
where $\mathbf{n}_{k} \in \mathbb{C}^{T\times 1}\sim \mathcal{CN}(\bm{0}, \sigma^{2})$ denotes the additive complex Gaussian noise, and $\textrm{Tr}(\mathbf{X}\mathbf{X}^{H}) = PTN$ with $P/\sigma^{2}$ representing
the signal-to-noise ratio (SNR). Due to the large number of antennas $N$ at the BS, it is difficult to recover $\mathbf{h}_{k}$ with high accuracy by employing traditional algorithms, e.g., least squares method. Thus, we employ the SBL technique to solve the channel estimation problem with limited training overhead.

For clarity, we drop the user’s index $k$ and denote the true AoDs as $\{ \phi_{j}, j = 1, 2, \cdots, J \}$, where $J=N_{c}N_{s}$. Let $ \{ \hat{\phi}_{\hat{j}} \}_{\hat{j}=1}^{\hat{J}} $ be a fixed sampling grid that uniformly covers the angular domain $[-\frac{\pi}{2}, \frac{\pi}{2}]$, where $\hat{J}$ denotes the number of grid points. When the grid is fine enough such that all the true AoDs $\{ \phi_{j}, j = 1, 2, \cdots, J \}$ lie on the grid, we can apply the model for $\mathbf{h}$:
\begin{equation}  \label{Dictionary}
\mathbf{h}=\mathbf{A}\mathbf{w},
\end{equation}
where $\mathbf{A}=[\mathbf{a}(\hat{\phi}_{1}), \mathbf{a}(\hat{\phi}_{2}), \cdots, \mathbf{a}(\hat{\phi}_{\hat{J}}) ]\in \mathbb{C}^{N \times \hat{J} }$, $\mathbf{a}(\phi)$ denotes the steering vector, and $\mathbf{w}\in \mathbb{C}^{\hat{J} \times 1}$ is a sparse vector whose non-zero elements correspond to the true directions at $\{ \phi_{j}, j = 1, 2, \cdots, J \}$. For instance, if the $\hat{j}$-th element of $\mathbf{w}$ is non-zero and its corresponding true direction is $\phi_{j}$, then we have $\phi_{j}=\hat{\phi}_{\hat{j}}$. 

The assumption that the true directions are located on the predefined spatial grid is impractical. To deal with such direction mismatch, we employ an off-grid model. In particular, if $\phi_{j}\notin \{ \hat{\phi}_{\hat{j}} \}_{\hat{j}=1}^{\hat{J}}$ and $\hat{\phi}_{n_{j}}, n_{j}\in \{1, 2, \cdots, \hat{J} \}$ is the nearest grid point to $\phi_{j}$, we can express $\phi_{j}$ as
\begin{equation} \label{Offgrid}
\phi_{j} = \hat{\phi}_{n_{j}} + \beta_{n_{j}},
\end{equation}
where $\beta_{n_{j}}$ denotes the off-grid gap. Based on \eqref{Offgrid}, we obtain $\mathbf{a}(\phi_{j}) = \mathbf{a}( \hat{\phi}_{n_{j}} + \beta_{n_{j}} )$. Then, $\mathbf{h}$ can be further rewritten as
\begin{equation}
\mathbf{h}=\mathbf{A}(\bm{\beta})\mathbf{w},
\end{equation}
where $\bm{\beta}=[\beta_{1}, \beta_{2}, \cdots, \beta_{\hat{J}} ]^{T}$, $\mathbf{A}(\bm{\beta})=[\mathbf{a}(\hat{\phi}_{1}+\beta_{1}), \mathbf{a}(\hat{\phi}_{2}+\beta_{2}), \cdots, \mathbf{a}(\hat{\phi}_{\hat{J}}+\beta_{\hat{J}}) ]$, and
\begin{equation}
\beta_{n_{j}} = \left\{
\begin{aligned}
&\phi_{j}-\hat{\phi}_{n_{j}}, & & l=1, 2, \cdots, L, \\
&0, & &\textrm{otherwise}. 
\end{aligned}
\right.
\end{equation}
With the off-grid basis, the model significantly reduces the direction mismatch since there always exists $\beta_{n_{j}}$ making \eqref{Offgrid} hold exactly. Then, the received signal is rewritten as
\begin{equation} \label{Offchannel}
\mathbf{y} = \mathbf{X}\mathbf{A}(\bm{\beta})\mathbf{w} + \mathbf{n} = \bm{\Phi}(\bm{\beta})\mathbf{w} + \mathbf{n},
\end{equation}
where $\bm{\Phi}(\bm{\beta}) \triangleq \mathbf{X}\mathbf{A}(\bm{\beta})$. 
Thus, the problem can be formulated as 
\begin{equation} \label{Problem}
\min\limits_{\mathbf{w}} \| \mathbf{w} \|_{0}, \quad \textrm{s.t.} \| \mathbf{y} - \mathbf{X} \mathbf{A}(\bm{\beta}) \mathbf{w} \|_{2} < \varsigma ,
\end{equation}
where $\varsigma$ is a constant that depends on $\|\mathbf{n}\|_{2}$. Since the coefficient vector $\bm{\beta}$ is unknown, the existing $l_{1}$-norm minimization methods cannot be employed to solve \eqref{Problem} directly. In the following, we model the distributions of the off-grid parameters and develop a SBL-based algorithm to jointly recover the sparse channel and refine the off-grid parameters.

\subsection{SBL Formulation}
Inspired from \cite{SBLChannel}, the distributions of parameters are modeled as follows. Under the assumption of circular symmetric complex Gaussian noise, we have
\begin{equation} \label{CSCGN}
p(\mathbf{y}|\mathbf{w}, \alpha, \boldsymbol{\beta})=\mathcal{CN} \left(\mathbf{y}|\mathbf{\Phi}(\boldsymbol{\beta}) \mathbf{w}, \alpha^{-1} \mathbf{I}\right),
\end{equation}
where $\alpha = \sigma_{n}^{-2}$ represents the noise precision. Since $\alpha$ is generally unknown, we model it as a Gamma hyperprior $p(\alpha) = \Gamma(\alpha; 1+a_{1}, b_{1})$, where we set $a_{1}, b_{1} \rightarrow 0$ as in \cite{SBLSense} to acquire a broad hyperprior. Then, we assume a non-informative i.i.d. uniform prior
for the elements of $\bm{\beta}$, 
\begin{equation}
p(\boldsymbol{\beta}) = \prod_{j=1}^{\hat{J}} U(\beta_{j};-\frac{\pi}{2}, \frac{\pi}{2}).
\end{equation}
Based on the sparse Bayesian model \cite{SBLSense}, we assign a Gaussian prior distribution with a distinct precision $\gamma_{i}$ for each element of $\mathbf{w}$.  Letting $\boldsymbol{\gamma} = [\gamma_{1}, \gamma_{2}, \cdots, \gamma_{\hat{J}} ]^{T}$ , we have
\begin{equation} \label{GaussPrior}
p(\mathbf{w}|\boldsymbol{\gamma}) = \mathcal{CN}\left(\mathbf{w}|\bm{0}, \textrm{diag}(\boldsymbol{\gamma}^{-1})\right).
\end{equation}
Then, we model the elements of $\boldsymbol{\gamma}$ as i.i.d. Gamma distributions,
\begin{equation}
p(\boldsymbol{\gamma}) = \prod_{j=1}^{\hat{J}}\Gamma (\gamma_{j};1+a_{2},b_{2}).
\end{equation}
Thus, the two-stage hierarchical prior gives
\begin{equation}
\begin{aligned}
p(\mathbf{w}) \!=\! \int_{0}^{\infty} \!\! p(\mathbf{w}|\boldsymbol{\gamma}) p(\boldsymbol{\gamma}) d \boldsymbol{\gamma} \!
\propto \! \prod_{i=1}^{\hat{J}} \! \left(  b_{2}+\left|w_{i}\right|^{2} \right)^{-(  a_{2}+\frac{3}{2} )} \!, 
\end{aligned}
\end{equation}
which encourages sparsity due to the heavy tails and sharp peak at $0$ with a small $b_{2}$ \cite{SBLSense}. 

The precision $\gamma_{j}$ in \eqref{GaussPrior} indicates the support of $\mathbf{w}$. For example, when $\gamma_{j}$ is large, the $j$-th element of $\mathbf{w}$ tends to be $0$. Otherwise, the value of the $l$-th element is significant. Therefore, once we obtain the precision vector $\boldsymbol{\gamma}$ and the off-grid gap $\boldsymbol{\beta}$, the estimated channel is given by
\begin{equation} \label{ChannelEsti}
\hat{\mathbf{h}} = \mathbf{A}_{\Omega}(\boldsymbol{\beta}) ( \bm{\Phi}_{\Omega}(\boldsymbol{\beta}) )^{\dagger} \mathbf{y},
\end{equation}
where $\Omega \triangleq \textrm{supp}(\mathbf{w})$. Hence, we aim to find the optimal $\boldsymbol{\beta}$ and $\boldsymbol{\gamma}$. As the noise precision $\alpha$ is unknown, we find the optimal values $\alpha^{\star}$, $\boldsymbol{\gamma}^{\star}$, and $\boldsymbol{\beta}^{\star}$ by maximizing the posteriori $p(\alpha, \boldsymbol{\gamma}, \boldsymbol{\beta}|\mathbf{y})$,
or equivalently,
\begin{equation} \label{OriginalObj}
(\alpha^{\star}, \boldsymbol{\gamma}^{\star}, \boldsymbol{\beta}^{\star}) = \arg \max\limits_{\alpha, \boldsymbol{\gamma}, \boldsymbol{\beta}} \ln p(\mathbf{y}, \alpha, \boldsymbol{\gamma}, \boldsymbol{\beta}).
\end{equation}
Note that the objective \eqref{OriginalObj} is a high-dimensional non-convex function. It is difficult to directly employ the gradient ascent method, since it has a slow convergence speed and may find unsatisfactory local optimum. In addition, the gradient of the original objective function \eqref{OriginalObj} has no closed-form expression.
To address these issues, the SBL-based algorithm is developed based on the SBL formulation and the framework of block MM algorithm. 
In the following, we show the detailed procedures of the SBL-based algorithm, which finds a stationary point of \eqref{OriginalObj} \cite{SBLChannel}. 

\subsection{The Proposed SBL-Based Algorithm} 

\subsubsection{Framework of Block MM Algorithm}
The block MM algorithm can efficiently solve the non-convex problem and accelerate the convergence speed \cite{MMAlgorithm}. It aims to iteratively construct a continuous surrogate function for the original objective function $\ln p(\mathbf{y}, \alpha, \boldsymbol{\gamma}, \boldsymbol{\beta})$, and then alternately maximize the surrogate function w.r.t. $\alpha$, $\boldsymbol{\gamma}$, and $\boldsymbol{\beta}$ as 
\begin{subequations} 
\begin{eqnarray}
& & \!\!\!\!\!\!\!\!\!\!\!\!\!\!\!\!\!\!\!\!\!\!\!
\alpha^{(t+1)} \!=\! \arg \max\limits_{\alpha} \mathcal{G} \! \left( \! \alpha, \boldsymbol{\gamma}^{(t)}, \boldsymbol{\beta}^{(t)} | \alpha^{(t)}, \boldsymbol{\gamma}^{(t)}, \boldsymbol{\beta}^{(t)} \! \right), \label{UpdateRule1} \\
& & \!\!\!\!\!\!\!\!\!\!\!\!\!\!\!\!\!\!\!\!\!\!\! \boldsymbol{\gamma}^{(t+1)} \!=\! \arg \max\limits_{\boldsymbol{\gamma}} \mathcal{G} \! \left( \! \alpha^{(t+1)}, \boldsymbol{\gamma}, \boldsymbol{\beta}^{(t)} | \alpha^{(t+1)}, \boldsymbol{\gamma}^{(t)}, \boldsymbol{\beta}^{(t)} \! \right), \label{UpdateRule2} \\
& & \!\!\!\!\!\!\!\!\!\!\!\!\!\!\!\!\!\!\!\!\!\!\! \boldsymbol{\beta}^{(t+1)} \!=\! \arg \max\limits_{\boldsymbol{\beta}} \mathcal{G} \! \left( \! \alpha^{(t+1)} \!,\! \boldsymbol{\gamma}^{(t+1)}\!,\! \boldsymbol{\beta} | \alpha^{(t+1)}\!,\! \boldsymbol{\gamma}^{(t+1)}\!,\! \boldsymbol{\beta}^{(t)} \! \right)\!,\! \label{UpdateRule3}
\end{eqnarray}
\end{subequations}
where $(\cdot)^{(t)}$ stands for the $t$-th iteration and $\mathcal{G}(\alpha, \boldsymbol{\gamma}, \boldsymbol{\beta} | \alpha^{(t)}, \boldsymbol{\gamma}^{(t)}, \boldsymbol{\beta}^{(t)})$ is the surrogate function constructed at the fixed point $(\alpha^{(t)}, \boldsymbol{\gamma}^{(t)}, \boldsymbol{\beta}^{(t)})$ that satisfies the properties: (i) It is the lower bound of $\ln p(\mathbf{y}, \alpha, \boldsymbol{\gamma}, \boldsymbol{\beta})$; (ii) Its value and partial derivative  w.r.t. $\alpha$, $\boldsymbol{\gamma}$, and $\boldsymbol{\beta}$ equal to those of $\ln p(\mathbf{y}, \alpha, \boldsymbol{\gamma}, \boldsymbol{\beta})$ at $(\alpha^{(t)}, \boldsymbol{\gamma}^{(t)}, \boldsymbol{\beta}^{(t)})$. 
The update rules \eqref{UpdateRule1}-\eqref{UpdateRule3} guarantee the convergence of the block MM algorithm \cite{MMAlgorithm}.

\subsubsection{Surrogate Function}
Inspired by the expectation-maximization (EM) algorithm \cite{SBLChannel, SBLSense,MMAlgorithm}, we construct the surrogate function at fixed point $(\alpha^{(t)}, \boldsymbol{\gamma}^{(t)}, \boldsymbol{\beta}^{(t)})$ as
\begin{equation}
\begin{aligned} 
& \mathcal{G}(\alpha, \boldsymbol{\gamma}, \boldsymbol{\beta} | \alpha^{(t)}, \boldsymbol{\gamma}^{(t)}, \boldsymbol{\beta}^{(t)}) \\
& \!\!\!\!\!\!\!\!\! = \int p(\mathbf{w} | \mathbf{y}, \alpha^{(t)}, \boldsymbol{\gamma}^{(t)}, \boldsymbol{\beta}^{(t)})  \ln \frac{p(\mathbf{w}, \mathbf{y}, \alpha, \boldsymbol{\gamma}, \boldsymbol{\beta})}{p(\mathbf{w} | \mathbf{y}, \alpha^{(t)}, \boldsymbol{\gamma}^{(t)}, \boldsymbol{\beta}^{(t)})} d \mathbf{w}.
\end{aligned} 
\end{equation}
According to \eqref{CSCGN} and \eqref{GaussPrior}, $p(\mathbf{w}|\mathbf{y}, \alpha, \boldsymbol{\gamma}, \boldsymbol{\beta})$ is complex Gaussian
\begin{equation}
p(\mathbf{w}|\mathbf{y}, \alpha, \boldsymbol{\gamma}, \boldsymbol{\beta}) = \mathcal{CN} ( \mathbf{w}|\boldsymbol{\mu}(\alpha, \boldsymbol{\gamma}, \boldsymbol{\beta}), \boldsymbol{\Sigma}(\alpha, \boldsymbol{\gamma}, \boldsymbol{\beta}) ),
\end{equation}
where
\begin{equation} 
\begin{aligned}
\boldsymbol{\mu}(\alpha, \boldsymbol{\gamma}, \boldsymbol{\beta}) &= \alpha \boldsymbol{\Sigma}(\alpha, \boldsymbol{\gamma}, \boldsymbol{\beta}) \mathbf{\Phi}^{H}(\boldsymbol{\beta})\mathbf{y}, \\
\boldsymbol{\Sigma}(\alpha, \boldsymbol{\gamma}, \boldsymbol{\beta}) &= ( \alpha \mathbf{\Phi}^{H}(\boldsymbol{\beta}) \mathbf{\Phi}(\boldsymbol{\beta}) + \textrm{diag}(\boldsymbol{\gamma}) )^{-1}.
\end{aligned}
\end{equation}
In the following, we present the detailed update procedures for $\alpha$, $\boldsymbol{\gamma}$, and $\boldsymbol{\beta}$. 

\subsubsection{Solutions for $\alpha$}
Firstly, the maximization problem in \eqref{UpdateRule1} has the following closed-form solution
\begin{equation}  \label{UpdateAlpha}
\alpha^{(t+1)}=\frac{T+a}{b+\eta\left(\alpha^{(t)}, \boldsymbol{\gamma}^{(t)}, \boldsymbol{\beta}^{(t)}\right)},
\end{equation}
where
\begin{equation} \label{eta}
\eta(\alpha \!, \boldsymbol{\gamma} \!, \boldsymbol{\beta}) \!=\! \textrm{Tr} \! \left( \! \boldsymbol{\Phi}(\boldsymbol{\beta}) \boldsymbol{\Sigma}(\alpha \!, \boldsymbol{\gamma} \!, \boldsymbol{\beta}) \boldsymbol{\Phi}^{H}(\boldsymbol{\beta}) \! \right) \!+\! \|\mathbf{y} \!-\! \boldsymbol{\Phi}(\boldsymbol{\beta}) \boldsymbol{\mu}(\alpha \!, \boldsymbol{\gamma} \!, \boldsymbol{\beta})\|^{2}.
\end{equation}

\subsubsection{Solutions for $\boldsymbol{\gamma}$}
Secondly, the problem \eqref{UpdateRule2} has the following closed-form solution
\begin{equation} \label{UpdateGamma}
\gamma^{(t+1)}_{j}=\frac{a+1}{b+[ \boldsymbol{\Lambda} \left(\alpha^{(t+1)}, \boldsymbol{\gamma}^{(t)}, \boldsymbol{\beta}^{(t)}\right) ]_{jj}}, \forall j,
\end{equation}
where
\begin{equation}
\boldsymbol{\Lambda}(\alpha, \boldsymbol{\gamma}, \boldsymbol{\beta})=\boldsymbol{\Sigma}(\alpha, \boldsymbol{\gamma}, \boldsymbol{\beta})+\boldsymbol{\mu}(\alpha, \boldsymbol{\gamma}, \boldsymbol{\beta}) \boldsymbol{\mu}^{H}(\alpha, \boldsymbol{\gamma}, \boldsymbol{\beta}).
\end{equation}

\begin{figure*}[t]
\begin{centering}
\includegraphics[width=0.9\textwidth]{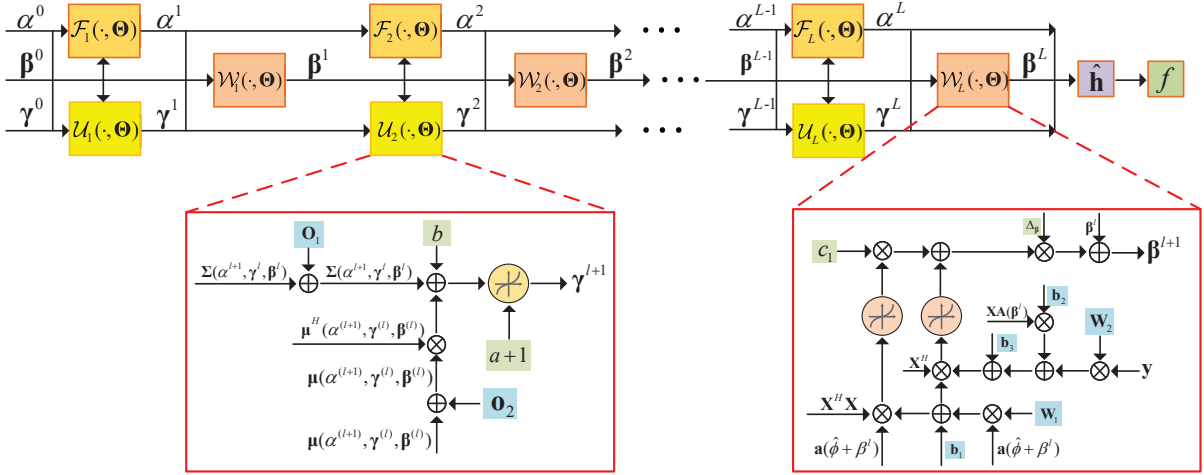}
\par\end{centering}
\caption{The deep-unfolding architecture of the SBL-based algorithm for channel estimation.}
\label{UnfoldArchitect}
\end{figure*}

\subsubsection{Solutions for $\boldsymbol{\beta}$}
Since problem \eqref{UpdateRule3} is non-convex, it is difficult to find its optimal solution. Thus, we employ gradient update on its objective function and obtain a one-step update for $\boldsymbol{\beta}$. The derivative of the
objective function in \eqref{UpdateRule3} w.r.t. $\boldsymbol{\beta}$ is computed as
\begin{equation}
\boldsymbol{\Xi}_{\boldsymbol{\beta}}^{(t)}=\left[\Xi^{(t)}\left(\beta_{1}\right), \Xi^{(t)}\left(\beta_{2}\right), \ldots, \Xi^{(t)}\left(\beta_{\hat{J}}\right)\right]^{T},
\end{equation}
with
\begin{equation}  \label{UpdateBeta}
\begin{aligned}
\Xi^{(t)}(\beta_{j}) &= 2 \Re \left\{ (\mathbf{a}^{\prime}(\hat{\phi}_{j}+\beta_{j}))^{H} \mathbf{X}^{H} \mathbf{X}(\mathbf{a}(\hat{\phi}_{j}+\beta_{j})) \right\}  \\
& \quad \cdot c_{1}^{(t)} + 2 \Re \left\{ (\mathbf{a}^{\prime}(\hat{\phi}_{j}+\beta_{j}))^{H} \mathbf{X}^{H} \mathbf{c}_{2}^{(t)} \right\},
\end{aligned}
\end{equation}
where 
\begin{subequations}  
\begin{eqnarray}
& & \!\!\!\!\!\!\! c_{1}^{(t)} \triangleq -\alpha^{(t+1)}(\chi_{jj}^{(t)}+|\mu_{j}^{(t)}|^{2}), \\ 
& & \!\!\!\!\!\!\! \mathbf{c}_{2}^{(t)} \triangleq \alpha^{(t+1)}\big( \mu_{j}^{(t)} \mathbf{y}_{-j}^{(t)}-\mathbf{X} \sum_{i \neq j} \chi_{ij}^{(t)} \mathbf{a}(\hat{\phi}_{i} + \beta_{i}) \big), \label{C2} \\
& & \!\!\!\!\!\!\! \mathbf{y}_{-j}^{(t)} \triangleq \mathbf{y}-\mathbf{X} \cdot \sum_{i \neq j}\big( \mu_{i}^{(t)} \cdot  \mathbf{a}(\hat{\phi}_{i}+\beta_{i}) \big), \\ 
& & \!\!\!\!\!\!\! \mathbf{a}^{\prime}(\hat{\phi}_{i}+\beta_{i}) \triangleq d \mathbf{a}(\hat{\phi}_{i}+\beta_{i}) / d \beta_{i}, 
\end{eqnarray}
\end{subequations}
$\mu_{j}^{(t)}$ and $\chi_{ij}^{(t)}$ denote the $j$-th element and the $(i,j)$-th element of $\boldsymbol{\mu}(\alpha^{(t+1)}, \boldsymbol{\gamma}^{(t+1)}, \boldsymbol{\beta}^{(t)})$ and $\boldsymbol{\Sigma}(\alpha^{(t+1)}, \boldsymbol{\gamma}^{(t+1)}, \boldsymbol{\beta}^{(t)})$, respectively.

The detailed derivation for \eqref{UpdateAlpha}-\eqref{UpdateBeta} can be obtained based on basic calculus in \cite{SBLChannel}. It is difficult to obtain the optimal solution for $\boldsymbol{\beta}$ and we only need to find a sub-optimal solution that increases the value of the objective function after each iteration. Hence, we employ the gradient descent method to update $\boldsymbol{\beta}$ as
\begin{equation}  \label{UpdateBeta1}
\boldsymbol{\beta}^{(t+1)} = \boldsymbol{\beta}^{(t)} + \Delta_{\boldsymbol{\beta}} \cdot \boldsymbol{\Xi}_{\boldsymbol{\beta}}^{(t)},
\end{equation}
where $\Delta_{\boldsymbol{\beta}}$ denotes the step-size. The detailed procedures of the proposed SBL-based algorithm are summarized in Algorithm \ref{SBLAlgo}.

\begin{algorithm}[t] \caption{The proposed SBL-based algorithm} \label{SBLAlgo}
\begin{algorithmic}[1]
\begin{small} 
\State \textbf{Input}: Pilot matrix $\mathbf{X}$, noise variance $\sigma^{2}$, transmission power $P$, pilot length $T$, and the precision of channel estimation $\delta$. The number of antennas $N$ and grid points $\hat{J}$.
\State \textbf{Initialize}: Variables $\alpha$, $\boldsymbol{\gamma}$, and $\boldsymbol{\beta}$. Set the iteration index $t=1$. 
\While{$\| \hat{\mathbf{h}}^{t} - \hat{\mathbf{h}}^{t-1} \|^{2} >\delta $}
\State Update $\alpha$ based on \eqref{UpdateAlpha};
\State Update $\boldsymbol{\gamma}$ based on \eqref{UpdateGamma};
\State Update $\boldsymbol{\beta}$ based on \eqref{UpdateBeta1};
\State Recover the channel $\hat{\mathbf{h}}^{t}$ based on \eqref{ChannelEsti};
\State t=t+1.
\EndWhile  
\end{small}
\end{algorithmic}
\end{algorithm}

\section{Deep-Unfolding for Channel Estimation} \label{DeepUnfold}

In this section, we unfold the proposed SBL-based algorithm into a layer-wise structure and provide the performance analysis.  

\subsection{Deep-Unfolding for the SBL-Based Algorithm}

As presented in Fig. \ref{UnfoldArchitect}, we unfold the proposed SBL-based algorithm into a layer-wise structure, where $\mathcal{F}_{l}(\cdot, \bm{\Theta})$, $\mathcal{U}_{l}(\cdot, \bm{\Theta})$, and $\mathcal{W}_{l}(\cdot, \bm{\Theta})$ denote the $l$-th layers that output $\alpha$, $\boldsymbol{\gamma}$, and $\boldsymbol{\beta}$, respectively. The trainable parameters $\bm{\Theta} = \{\bm{\Theta}_{1}, \bm{\Theta}_{2}\}$ can be divided into two categories, where we omit the index of $a$ and $b$ and the layer index $l$ for all the trainable parameters for clarity: (i) The hyper-parameters of the prior distributions and step size in the SBL-based algorithm that are difficult to determine, i.e., $\bm{\Theta}_{1} = \{a, b, c_{1}, \Delta_{\boldsymbol{\beta}} \}$; (ii) The introduced trainable parameters to replace the operations with high computational complexity and improve the performance, i.e., $\bm{\Theta}_{2} = \{ \mathbf{W}_{1}, \mathbf{W}_{2}, \mathbf{O}_{1}, \mathbf{o}_{2}, \mathbf{b}_{1}, \mathbf{b}_{2}, \mathbf{b}_{3} \} $.
Then, the deep-unfolding structure is expressed in \eqref{ArchiUnfold}.
\begin{figure*}
\begin{subequations}  \label{ArchiUnfold}
\begin{eqnarray}
& & \alpha^{(l+1)}=(T+a)\bigg( b + \textrm{Tr}\left(\boldsymbol{\Phi}(\boldsymbol{\beta}^{(l)}) \tilde{\boldsymbol{\Sigma}}(\alpha^{(l)}, \boldsymbol{\gamma}^{(l)}, \boldsymbol{\beta}^{(l)}) \boldsymbol{\Phi}^{H}(\boldsymbol{\beta}^{(l)})\right) +\|\mathbf{y}- \boldsymbol{\Phi}(\boldsymbol{\beta}^{(l)}) \tilde{\boldsymbol{\mu}}(\alpha^{(l)}, \boldsymbol{\gamma}^{(l)}, \boldsymbol{\beta}^{(l)})\|_{2}^{2} \bigg)^{-1},  \\
& & \gamma^{(l+1)}_{j}=(a+1) \bigg(b+[ \tilde{\boldsymbol{\Sigma}}(\alpha^{(l+1)}, \boldsymbol{\gamma}^{(l)}, \boldsymbol{\beta}^{(l)}) + \tilde{\boldsymbol{\mu}}(\alpha^{(l+1)}, \boldsymbol{\gamma}^{(l)}, \boldsymbol{\beta}^{(l)}) \tilde{\boldsymbol{\mu}}^{H}(\alpha^{(l+1)}, \boldsymbol{\gamma}^{(l)}, \boldsymbol{\beta}^{(l)}) ]_{jj} \bigg)^{-1}, \forall j,  \\
& & \beta^{(l+1)}_{j} = \beta^{(l)}_{j} + \Delta_{\beta_{j}}^{(l)} \cdot \bigg\{ \Re \bigg( (  \mathbf{W}_{1}\mathbf{a}(\hat{\phi}_{j}+\beta_{j}) + \mathbf{b}_{1} )^{H} \mathbf{X}^{H} \mathbf{X}(\mathbf{a}(\hat{\phi}_{j} + \beta_{j}))  \bigg) c_{1}^{(l)} \notag \\
& & \quad \quad \quad \quad \quad \quad \quad \quad \quad \quad \quad \quad + \Re \bigg( ( \mathbf{W}_{1}\mathbf{a}(\hat{\phi}_{j}+\beta_{j}) + \mathbf{b}_{1} )^{H} \mathbf{X}^{H} ( \mathbf{W}_{2}\mathbf{y} +\mathbf{X}\mathbf{A}(\bm{\beta})\mathbf{b}_{2}+\mathbf{b}_{3} )  \bigg)  \bigg\},  \forall j. 
\end{eqnarray}
\end{subequations}
\end{figure*}

We introduce the trainable parameters $\bm{\Theta}_{2}$ in the following aspects:
\begin{itemize}
\item Since $\boldsymbol{\Sigma}(\alpha, \boldsymbol{\gamma}, \boldsymbol{\beta}) $ and $\boldsymbol{\mu}(\alpha, \boldsymbol{\gamma}, \boldsymbol{\beta})$ tend to be imprecise, which cannot depict the statistics of $p(\mathbf{w}|\mathbf{y}, \alpha, \boldsymbol{\gamma}, \boldsymbol{\beta})$ accurately. To improve the precision, we introduce trainable parameters $\mathbf{O}_{1}$ and $\mathbf{o}_{2}$ to refine them as $\tilde{\boldsymbol{\Sigma}}(\alpha, \boldsymbol{\gamma}, \boldsymbol{\beta}) = \boldsymbol{\Sigma}(\alpha, \boldsymbol{\gamma}, \boldsymbol{\beta}) + \mathbf{O}_{1}$ and $\tilde{\boldsymbol{\mu}}(\alpha, \boldsymbol{\gamma}, \boldsymbol{\beta}) = \boldsymbol{\mu}(\alpha, \boldsymbol{\gamma}, \boldsymbol{\beta}) + \mathbf{o}_{2}$, respectively.	
	
\item Recall that $\mathbf{a}^{\prime}(\hat{\phi}_{i}+\beta_{i}) = d \mathbf{a}(\hat{\phi}_{i}+\beta_{i}) / d \beta_{i}$, the element of $\mathbf{a}(\hat{\phi}_{i}+\beta_{i})$ has the form $e^{-j2\pi \textrm{sin}(\hat{\phi}_{i}+\beta_{i}) }$, and its derivative is $-j2\pi \textrm{cos}(\hat{\phi}_{i}+\beta_{i}) e^{-j2\pi \textrm{sin}(\hat{\phi}_{i}+\beta_{i}) }$. Then, $\mathbf{a}^{\prime}(\hat{\phi}_{i}+\beta_{i})$ can be written as $\bm{\Phi}\mathbf{a}(\hat{\phi}_{i}+\beta_{i})$, where $\bm{\Phi}$ consists of $-j2\pi \textrm{cos}(\hat{\phi}_{i}+\beta_{i})$. Hence, we replace $\mathbf{a}^{\prime}(\hat{\phi}_{i}+\beta_{i})$ with $\mathbf{W}_{1}\mathbf{a}(\hat{\phi}_{i}+\beta_{i}) + \mathbf{b}_{1}$, where $\mathbf{W}_{1}$ and $\mathbf{b}_{1}$ are introduced trainable parameters.

\item Recall the expression of $\mathbf{c}_{2}$ in \eqref{C2} and $\mathbf{A}(\bm{\beta})=[\mathbf{a}(\hat{\phi}_{1}+\beta_{1}), \mathbf{a}(\hat{\phi}_{2}+\beta_{2}), \cdots, \mathbf{a}(\hat{\phi}_{\hat{J}}+\beta_{\hat{J}}) ]$. Then, $\mathbf{X} \sum_{i \neq j} \chi_{ij}^{(l)} \mathbf{a}(\hat{\phi}_{i} + \beta_{i})$ can be rewritten as $\mathbf{X}\mathbf{A}(\bm{\beta})\bm{\chi}$, where $\bm{\chi}$ consists of $\chi_{ij}$. Thus, we employ the structure  $\mathbf{W}_{2}\mathbf{y}+\mathbf{X}\mathbf{A}(\bm{\beta})\mathbf{b}_{2}+\mathbf{b}_{3}$ to replace $\mathbf{c}_{2}$, where $\mathbf{W}_{2}$, $\mathbf{b}_{2}$, and $\mathbf{b}_{3}$ are introduced trainable parameters.

\item The pilot matrix $\mathbf{X}$ can be treated as trainable parameters since the trained $\mathbf{X}$ better fits the current CSI statistics and achieves better accuracy with reduced pilot length.
\end{itemize}

Based on the optimized variables $\alpha$, $\boldsymbol{\gamma}$, and $\boldsymbol{\beta}$, \eqref{ChannelEsti} is employed to reconstruct the channel. Moreover, we apply the following NMSE as the loss function
\begin{equation}
\frac{1}{|\mathcal{M}|}\sum\limits_{m=1}^{|\mathcal{M}|}\dfrac{\| \hat{\mathbf{h}}_{m} - \mathbf{h}_{m} \|^{2} }{ \| \mathbf{h}_{m} \|^{2} },
\end{equation}
where $|\mathcal{M}|$ is the number of training dataset, $\hat{\mathbf{h}}_{m}$ is the estimation of the true channel $\mathbf{h}_{m}$ at the $m$-th sample. Furthermore, we provide the performance analysis of the proposed deep-unfolding NN in the following subsection.

\subsection{Performance Analysis of the Proposed Deep-Unfolding NN} \label{Analysis}
Generally, it is difficult to provide the general performance analysis of deep-unfolding NN due to the introduced trainable parameters and the differences between its architecture and that of the iterative algorithm. In this work, we propose the following theorem
\begin{theorem} \label{theorem1}
There exist the trainable parameters ensuring that the performance achieved by one layer of the deep-unfolding NN can approach that of two iterations of the optimization algorithm. 
\end{theorem}

The proof of Theorem \ref{theorem1} is provided in Appendix \ref{AppendixA}. This proof can be straightforwardly extended to demonstrate: There exist the trainable parameters ensuring that the performance achieved by one layer of the deep-unfolding NN can approach that of several iterations of the optimization algorithm. Thus, the proposed deep-unfolding NN achieves approaching performance of the SBL-based algorithm with reduced number of layers. Furthermore, the required number of layers varies from the channel samples and the adaptive depth is required to find the optimal number of layers for each CSI sample.  

\section{DDPG-Driven Deep-Unfolding} \label{DDPGChannel}

In this section, we design the DDPG-driven deep-unfolding for the SBL-based algorithm with adaptive depth and propose the halting score to control the channel reconstruction error. 

\subsection{DDPG-Driven Deep-Unfolding for SBL-Based Algorithm} \label{DDPGUnfo}

\begin{figure*}[t]
\begin{centering}
\includegraphics[width=0.8\textwidth]{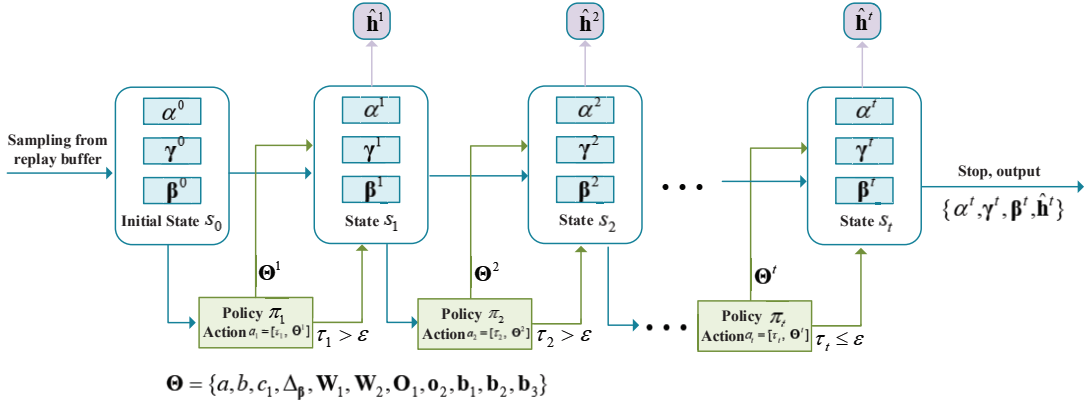}
\par\end{centering}
\caption{DDPG-driven deep-unfolding of the SBL-based algorithm for channel estimation.}
\label{DDPGUnfold}
\end{figure*}

The DDPG-driven deep-unfolding framework is employed to solve this problem, as shown in Fig. \ref{DDPGUnfold}. The MDP is formulated as below.

\begin{itemize}
\item State space: It consists of the optimization variable $\{\alpha, \boldsymbol{\gamma}, \boldsymbol{\beta}\}$, which includes the initialization $\{\alpha^{0}, \boldsymbol{\gamma}^{0}, \boldsymbol{\beta}^{0}\}$ and all intermedia results $\{\alpha^{l}, \boldsymbol{\gamma}^{l}, \boldsymbol{\beta}^{l}, \forall l\}$ in the deep-unfolding process, i.e., the output of each layer of the deep-unfolding NN. In particular, the state at the $t$-th time step is $s_{t}\triangleq [\alpha^{t}, \boldsymbol{\gamma}^{t}, \boldsymbol{\beta}^{t}]$.

\item Action space: It is composed of the halting indicator $\tau$ and the trainable parameters $\{ \bm{\Theta}^{l}, \forall l \}$ in each layer. Specifically, the action at the $t$-th time step is $a_{t}\triangleq [\tau_{t}, \bm{\Theta}^{t}]$. The role of $\tau_{t}\in [0,1]$ is to determine whether to halt the running of deep-unfolding NN at the current layer. The halting indicator $\tau_{t}$ is learned directly by DDPG and it moves forward to the next time step if $\tau_{t} > \varepsilon$, where $\varepsilon$ is a hyper-parameter. Otherwise the DDPG-driven deep-unfolding would be halted to output the current state as final results.   

\item State transition: Based on the observed state $s_{t}=[\alpha^{t}, \boldsymbol{\gamma}^{t}, \boldsymbol{\beta}^{t}]$ and the selected action $a_{t}=[\tau_{t}, \bm{\Theta}^{t}]$, the state transition is composed of one or several layers of deep-unfolding NN designed in \eqref{ArchiUnfold} if $\tau_{t} > \varepsilon$. For example, the state transition is defined as $[\alpha^{l}, \boldsymbol{\gamma}^{l}, \boldsymbol{\beta}^{l}]\rightarrow [\alpha^{l+1}, \boldsymbol{\gamma}^{l+1}, \boldsymbol{\beta}^{l+1}]$ when it contains one layer of deep-unfolding NN. Otherwise, when $\tau_{t} \leq \varepsilon$, the DDPG-driven deep-unfolding would treat the current state as the final state and output it.

\item Reward: The reward is designed as the decrement of NMSE performance between the former and current time step
\begin{equation} \label{rewardfunc}
r_{t} = \dfrac{\| \hat{\mathbf{h}}^{t-1} - \mathbf{h} \|^{2} }{ \| \mathbf{h} \|^{2} } - \dfrac{\| \hat{\mathbf{h}}^{t} - \mathbf{h} \|^{2} }{ \| \mathbf{h} \|^{2} } - \eta, 
\end{equation}
where $\hat{\mathbf{h}}^{t}$ is the estimation of true channel $\mathbf{h}$ at the $t$-th time step. A higher reward is received when the policy results in higher performance improvement. In addition, $\eta$ is a constant and it penalizes the policy as it does not halt at time step $t$. A negative reward will be given if the performance improvement cannot exceed the penalty $\eta$, thus forcing the policy to stop with diminished reward.
\end{itemize}

Based on the performance analysis of DDPG in \cite{DDPG0,TuningFree,DDPG1,DDPG2} and deep-unfolding NN in Section \ref{Analysis}, we can conclude that the proposed DDPG-driven deep-unfolding can achieve approaching performance of the SBL-based algorithm for each CSI sample with adaptive depth.

\subsection{Improvement of DDPG-Driven Deep-Unfolding}

The design of halting indicator $\tau_{t}$ mentioned above is a general method to indicate when to halt. As for this channel estimation problem, it cannot control the channel reconstruction error $\| \mathbf{h} - \hat{\mathbf{h}}^{t} \|^{2}$ accurately and the hyper-parameter $\epsilon$ is hard to determine. To address these issues, we propose a more accurate and effective method to design $\tau_{t}$ for this problem. 

\subsubsection{Halting Score} \label{HaltScore}

We introduce the halting scores \cite{AdaptDepth} $L_{t}\in [0, 1]$ in each layer $t$ to indicate whether to stop. Ideally, the halting score should be related to channel reconstruction error $\| \mathbf{h} - \hat{\mathbf{h}}^{t} \|^{2}$, but we do not know the true $\mathbf{h}$. To approximate the channel reconstruction error, we consider $\| \mathbf{y}-\mathbf{X} \hat{\mathbf{h}}^{t} \|^{2}$, which is the stopping criterion in many algorithms \cite{AdaptDepth,SBLChannel}. 
Here we use $\| \mathbf{Q} ( \mathbf{y}-\mathbf{X} \hat{\mathbf{h}}^{t} ) \|^{2}$ as an approximation of channel reconstruction error at the $t$-th layer, where $\mathbf{Q}$ is a linear mapping for the residual $(\mathbf{y}-\mathbf{X} \hat{\mathbf{h}}^{t}) = \mathbf{X}(\mathbf{h} - \hat{\mathbf{h}}^{t})$. Since it is difficult to determine the mapping matrix $\mathbf{Q}$ directly, we aim to learn $\mathbf{Q}$ together with the trainable parameters in DDPG. Thus, the halting score function is designed as
\begin{equation} \label{HaltScore1}
L_{t}=\sigma ( p_{1} \| \mathbf{Q} ( \mathbf{y}-\mathbf{X} \hat{\mathbf{h}}^{t} ) \|^{2} + p_{2} ),
\end{equation}
where $p_{1}>0$, $p_{2}$, and $\mathbf{Q}$ are trainable parameters, $\sigma(\cdot)$ is the sigmoid function which returns the value from $0$ to $1$, and $p_{1}>0$ ensures that the halting score decreases with the residual $\| \mathbf{Q} ( \mathbf{y}-\mathbf{X} \hat{\mathbf{h}}^{t} ) \|^{2}$.
Based on \eqref{HaltScore1}, we further employ a DNN with $r$ fully-connected layers to learn the halting score as 
\begin{equation} \label{HaltScore2}
L_{t}=\sigma_{r} ( \mathbf{q}_{r}^{T} ( \cdots \sigma_{2}( \mathbf{Q}_{2}\sigma_{1} ( \mathbf{Q}_{1} ( \mathbf{y}-\mathbf{X} \hat{\mathbf{h}}^{t} )  + \mathbf{p}_{1} )+\mathbf{p}_{2}) \cdots ) + p_{r}),
\end{equation}
where $\mathbf{p}_{i}, \mathbf{Q}_{i}, \forall i=1,2,\cdots, r$ are trainable parameters, the trainable vector $\mathbf{q}_{r}^{T}$ and scalar $p_{r}$ are introduced to ensure that the output of this DNN is a scalar. The input and output of DNN are residual $(\mathbf{y}-\mathbf{X}\hat{\mathbf{h}}^{t})$ and halting score $L_{t}$, respectively.

With a properly designed cost function, the trainable parameters could be trained to better fit the error distribution, which approximates $\| \mathbf{h} - \hat{\mathbf{h}}^{t} \|^{2}$ more accurately. Then, we design a differentiable cost function as
\begin{equation} \label{CostFunc}
\mathcal{L}(\bm{\theta}) = \sum\limits_{t}  \frac{\| \mathbf{h} - \hat{\mathbf{h}}^{t} \|^{2}}{L_{t}} + \rho L_{t},
\end{equation}
where $\rho \geq 0$ is a regularization parameter, $\bm{\theta}$ denotes the set of trainable parameters of this DNN, and $\mathbf{h}$ denotes the true CSI, which is known in the training stage. The summation is from the first layer to the current $t$-th layer. 
Different from the iterative algorithms and the existing deep-unfolding NN where intermediate results have no explicit contribution to the cost function, the reconstructed $\hat{\mathbf{h}}^{t}$ and halting scores $L_{t}$ of all layers contribute to the cost function in \eqref{CostFunc}.
By letting the derivative of \eqref{CostFunc} w.r.t. $L_{t}$ be $0$, the learned optimal halting score is given by
\begin{equation} \label{Derivative}
L_{t}= \frac{\| \mathbf{h} - \hat{\mathbf{h}}^{t} \|}{\sqrt{\rho}}.
\end{equation}
Thus, a well-trained DNN generates halting scores proportional to the channel reconstruction error. The number of layers is determined by the first index where the halting score is smaller than $\epsilon$,
\begin{equation} \label{StopCondition}
T_{s}=\min\{t: L_{t} \leq \epsilon \}, 
\end{equation}
where $\epsilon$ is a pre-determined small constant. Changing the value of the halting constant $\epsilon$ generally results in a different number of executed layers and a varying channel reconstruction error. 
Note that $L_{t}$ is related to the channel reconstruction error and regularization parameter $\rho$ and the derivative in \eqref{CostFunc} comes to be $0$ at $L_{t}= \frac{\| \mathbf{h} - \hat{\mathbf{h}}^{t} \|}{\sqrt{\rho}}$.
Thus, we can tune the hyper-parameters $\epsilon$ and $\rho$ to realize different channel reconstruction error.
For example, if we expect to terminate at the layer with $\| \mathbf{h} - \hat{\mathbf{h}}^{t} \|^{2}=0.01$, we can select $\rho =1$ and $\epsilon = 0.1$.    

\subsubsection{Improvement of the DDPG-Driven Deep-Unfolding with Halting Score}  \label{ImproveDDPG}

We employ the halting score $L_{t}$ proposed in Section \ref{HaltScore} as $\tau_{t}$ and the MDP designed in Section \ref{DDPGUnfo} should be modified in the following aspects:

\begin{itemize}
\item The input of the DNN in \eqref{HaltScore2} is residual $(\mathbf{y}-\mathbf{X}\hat{\mathbf{h}}^{t})$ and the input of the actor network is the state $s_{t}$. Thus, $(\mathbf{y}-\mathbf{X}\hat{\mathbf{h}}^{t})$ should be designed as part of the state, i.e., $s_{t}\triangleq [\alpha^{t}, \boldsymbol{\gamma}^{t}, \boldsymbol{\beta}^{t}, (\mathbf{y}-\mathbf{X}\hat{\mathbf{h}}^{t})]$ and input together with $[\alpha^{t}, \boldsymbol{\gamma}^{t}, \boldsymbol{\beta}^{t}]$ into the actor network in DDPG.

\item The output of the DNN in \eqref{HaltScore2} is halting score $L_{t}$ and $L_{t}$ is part of the action $a_{t}\triangleq [L_{t}, \bm{\Theta}^{t}]$, where $a_{t}$ is the output of the actor network in DDPG. Thus, the DNN in \eqref{HaltScore2} can be treated as a sub-network of the actor network that outputs $L_{t}$ to indicate whether to terminate.

\item The reward is modified as the weighted sum of the reward function in \eqref{rewardfunc} and the cost function in \eqref{CostFunc}. In particular, \eqref{rewardfunc} ensures that the channel reconstruction error significantly decreases in each layer, while \eqref{CostFunc} is designed to control the channel reconstruction error  and the number of layers by tuning the hyper-parameters $\epsilon$ and $\rho$. 
\end{itemize} 

\subsection{Extension to DDPG-Driven Black-Box DNN}
The DDPG-driven deep-unfolding framework can also be employed to realize adaptive depth of the general black-box DNN. Then, we take the DNN with fully-connected layers as an example and the other architectures, e.g., CNN and ResNet, can be designed similarly. The fully-connected layer can be expressed as
\begin{equation} \label{ArchiDNN}
\mathbf{y}^{l+1}=\varphi (\mathbf{W}^{l}\mathbf{y}^{l}+\mathbf{b}^{l}),
\end{equation} 
where $\mathbf{y}^{l}$ denotes the output of the $l$-th layer, $\varphi$ is the non-linear function, e.g., sigmoid, $\mathbf{W}^{l}$ and $\mathbf{b}^{l}$ are the trainable weight and offset of the $l$-th layer, respectively. Its MDP can be formulated as below, where the design of reward is similar to Section \ref{DDPGUnfo}.

\begin{itemize}
\item State: The state at the $t$-th time step is $s_{t}\triangleq \mathbf{y}^{t}$, which is the output of the $(t-1)$-th layer in black-box DNN.

\item Action: The action at the $t$-th time step is $a_{t}\triangleq [\tau_{t}, \mathbf{W}^{t}, \mathbf{b}^{t}]$. The role of $\tau_{t}\in [0,1]$ is to determine whether to halt the running of DNN at the current layer. The DDPG continues to execute the next time step if $\tau_{t} > \varepsilon$. 

\item State transition: Based on the observed state $s_{t}=\mathbf{y}^{t}$ and the selected action $a_{t}=[\tau_{t}, \mathbf{W}^{t}, \mathbf{b}^{t}]$, the state transition is composed of one or several layers of the DNN in \eqref{ArchiDNN} if $\tau_{t} > \varepsilon$. For example, the state transition is defined as $\mathbf{y}^{l} \rightarrow \mathbf{y}^{l+1}$ when it contains one layer of the DNN.   
\end{itemize}

As for this channel estimation problem, we employ the black-box DNN with fully-connected layers \eqref{ArchiDNN} to learn $\alpha$, $\boldsymbol{\gamma}$, and $\boldsymbol{\beta}$. Thus, the state of DDPG is defined as $s_{t} = [\alpha, \boldsymbol{\gamma}, \boldsymbol{\beta}]$. Finally, the DDPG-driven black-box DNN outputs $[\alpha, \boldsymbol{\gamma}, \boldsymbol{\beta}]$ and the estimated channel can be obtained according to \eqref{ChannelEsti}.
Furthermore, we could employ the halting score $L_{t}$ as $\tau_{t}$ and the MDP is modified in a similar way as Section \ref{ImproveDDPG}.

\section{Simulation Results} \label{Simulation}
In this section, we verify the effectiveness of the proposed DDPG-driven deep-unfolding with adaptive layers.

\subsection{Simulation Setup}
We consider the scenario where the BS is equipped with a ULA with $N=128$ antennas and it sends the training pilot symbols with $T=60$ and $\textrm{SNR} = 20$ dB. We employ the 3GPP spatial channel model (SCM) to generate the channel coefficients for an urban microcell. 
The different number of rays $J$ leads to various sparsity levels of the channel, which requires different number of layers (iterations). For the richness of samples, the dataset consists of the channel with the number of rays ranging from $6$ to $20$. 
We provide the simulation results of the following algorithms:
\begin{itemize}
\item DDPG Unfolding/Adaptive: The proposed DDPG-driven deep-unfolding with adaptive layers;
\item Unfolding/Fixed: The proposed deep-unfolding of the SBL-based algorithm with an off-grid basis and the fixed number of layers;
\item DDPG Black-box/Adaptive: The proposed black-box DNN with adaptive layers;
\item Black-box/Fixed: The black-box DNN with fixed number of layers \cite{DNNchannel2};
\item SBL Off-grid: The SBL-based algorithm with the off-grid basis \cite{SBLChannel};
\item Standard SBL: The standard SBL method \cite{Dictionary1} with the dictionary $\mathbf{A}$ defined in \eqref{Dictionary};
\item Two-stage CS: The two-stage compressed sensing (CS) \cite{TwoStage}.
\end{itemize}

In the simulation, the iterative optimization algorithms, i.e., SBL Off-grid, Standard SBL, and Two-stage CS, conduct until convergence. We select an optimal number of layers $L_{opt}$ for Unfolding/Fixed and Black-box/Fixed, where they have satisfactory performance with $L_{opt}$ and the performance cannot be further increased with the increase of $L_{opt}$. For fairness, we tune the hyper-parameter $\epsilon$ in \eqref{StopCondition} to ensure that the averaged number of layers of the whole dataset in DDPG Unfolding/Adaptive is the same as that of the Unfolding/Fixed.  

\subsection{NMSE Performance}

\begin{figure}[t]
\begin{centering}
\includegraphics[width=0.39\textwidth]{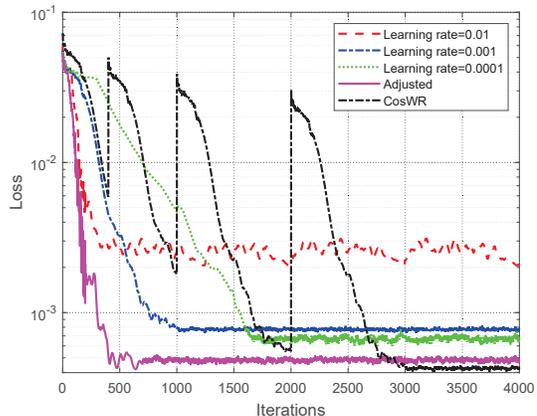}
\par\end{centering}
\caption{Convergence performance of NMSE with different learning rates.}
\label{LearningRate}
\end{figure}

Fig. \ref{LearningRate} presents the convergence performance of NMSE with different learning rates. It can be seen that the smaller learning rate achieves better NMSE performance but leads to a smaller convergence speed. Moreover, the adjusted learning rate that gradually decreases from $10^{-2}$ to $10^{-4}$ has satisfactory performance with fast convergence speed. The learning rate with cosine annealing and warm restart \cite{CosWR}, i.e., CosWR, achieves the best performance, but with relatively slower convergence speed compared to the adjusted learning rate. It has periodic oscillation due to the warm restart. 	

\begin{figure}[t]
\begin{centering}
\includegraphics[width=0.38\textwidth]{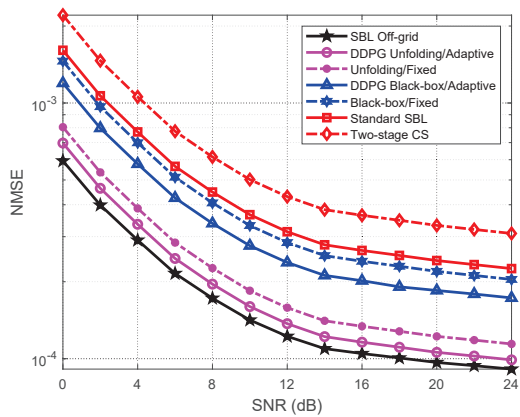}
\par\end{centering}
\caption{NMSE performance versus the SNR.}
\label{SNR}
\end{figure}

Fig. \ref{SNR} presents the NMSE performance of the proposed DDPG-driven deep-unfolding with adaptive layers and the benchmarks with different values of SNR. 
It is readily seen that the NMSE achieved by all algorithms decreases with SNR. The proposed DDPG Unfolding/Adaptive outperforms the Unfolding/Fixed, DDPG Black-box/Adaptive, Black-box/Fixed, Standard SBL, and Two-stage CS, where the gap increases with SNR. The SBL Off-grid achieves the best performance since it is an efficient iterative algorithm which takes the off-grid parameters into consideration and is guaranteed to find a local optimum. In addition, our proposed DDPG Unfolding/Adaptive achieves the NMSE performance approaching that of the SBL Off-grid. Thus, the proposed DDPG-driven deep-unfolding with adaptive layers is indeed an efficient framework for solving the channel estimation problem.

\begin{figure}[t]
\begin{centering}
\includegraphics[width=0.39\textwidth]{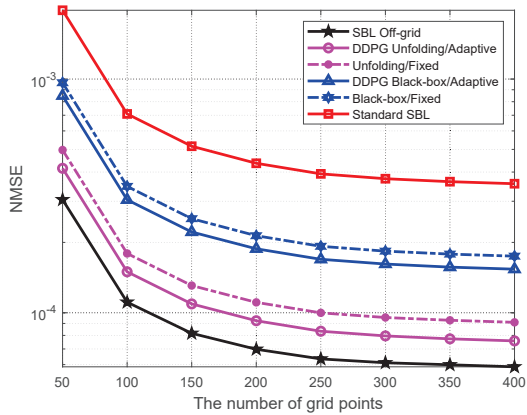}
\par\end{centering}
\caption{NMSE performance versus the number of grid points $\hat{J}$.}
\label{GridPoint}
\end{figure}

Fig. \ref{GridPoint} presents the NMSE performance versus the number of grid points $\hat{J}$.
The NMSE achieved by all algorithms decreases as $\hat{J}$ increases, since the dense grid leads to the precise estimation of AoDs. Furthermore, our proposed DDPG Unfolding/Adaptive approaches the performance achieved by the SBL Off-grid and always outperforms the others. In particular, the gap increases with $\hat{J}$ since the SBL Off-grid and DDPG Unfolding/Adaptive can obtain a more precise estimation of off-grid parameters and AoDs.
Moreover, we can see that the proposed SBL Off-grid significantly achieves better performance than the Standard SBL. It is mainly because: (i) The solution of Standard SBL is not exactly sparse, and it has performance loss due to the direction mismatch and energy leakage; (ii) The SBL Off-grid algorithm significantly improves the sparsity and accuracy of CSI representation, and the direction mismatch can be almost eliminated.

\begin{figure}[t]
\begin{centering}
\includegraphics[width=0.39\textwidth]{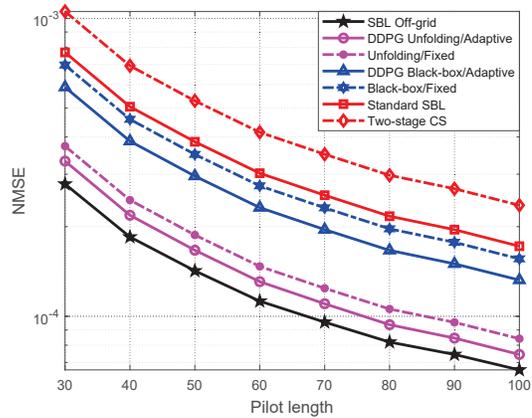}
\par\end{centering}
\caption{NMSE performance versus pilot length $T$.}
\label{PilotLength}
\end{figure}

Fig. \ref{PilotLength} shows the NMSE performance versus pilot length $T$. It is readily seen that the NMSE achieved by all schemes decreases as $T$ increases. The Standard SBL and Two-stage CS provide the worst performance, since they ignore the off-grid parameters. In addition, both the Black-box/Adaptive and Black-box/Fixed improve the NMSE performance, but the improvement is not so significant. Though the black-box methods consider the off-grid parameters, they do not employ the structure of iterative algorithms. In comparison, the proposed DDPG Unfolding/Adaptive unfolds the SBL-based algorithm with adaptive layers. It outperforms the benchmarks with the same length of pilots and approaches that of the SBL Off-grid. 

\subsection{Required Number of Layers}

\begin{figure}[t]
\begin{centering}
\includegraphics[width=0.39\textwidth]{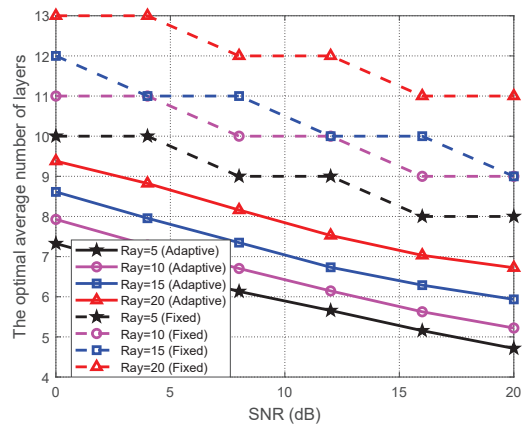}
\par\end{centering}
\caption{The required number of layers for different numbers of rays and SNR.}
\label{AverageLayers}
\end{figure}

Fig. \ref{AverageLayers} presents the required number of layers for different numbers of rays $J$ and SNR. The dotted lines denote the optimal number of layers of the deep-unfolding with fixed depth. In comparison, the full lines represent the required number of layers of the DDPG-driven deep-unfolding with adaptive depth to achieve nearly the same NMSE performance. It is readily seen that the proposed DDPG-driven deep-unfolding achieves approaching performance to the fixed-depth deep-unfolding with around $30\%$ reduced number of layers. Furthermore, the required number of layers increases with $J$, where a larger $J$ generally leads to higher sparsity level of channel. Besides, the required number of layers decreases as the increment of SNR. It is mainly because the larger the SNR is, the more accurate the channel estimation will be.

\begin{figure*}[!t]
\centering
\subfloat[]{\centering \scalebox{0.5}{\includegraphics{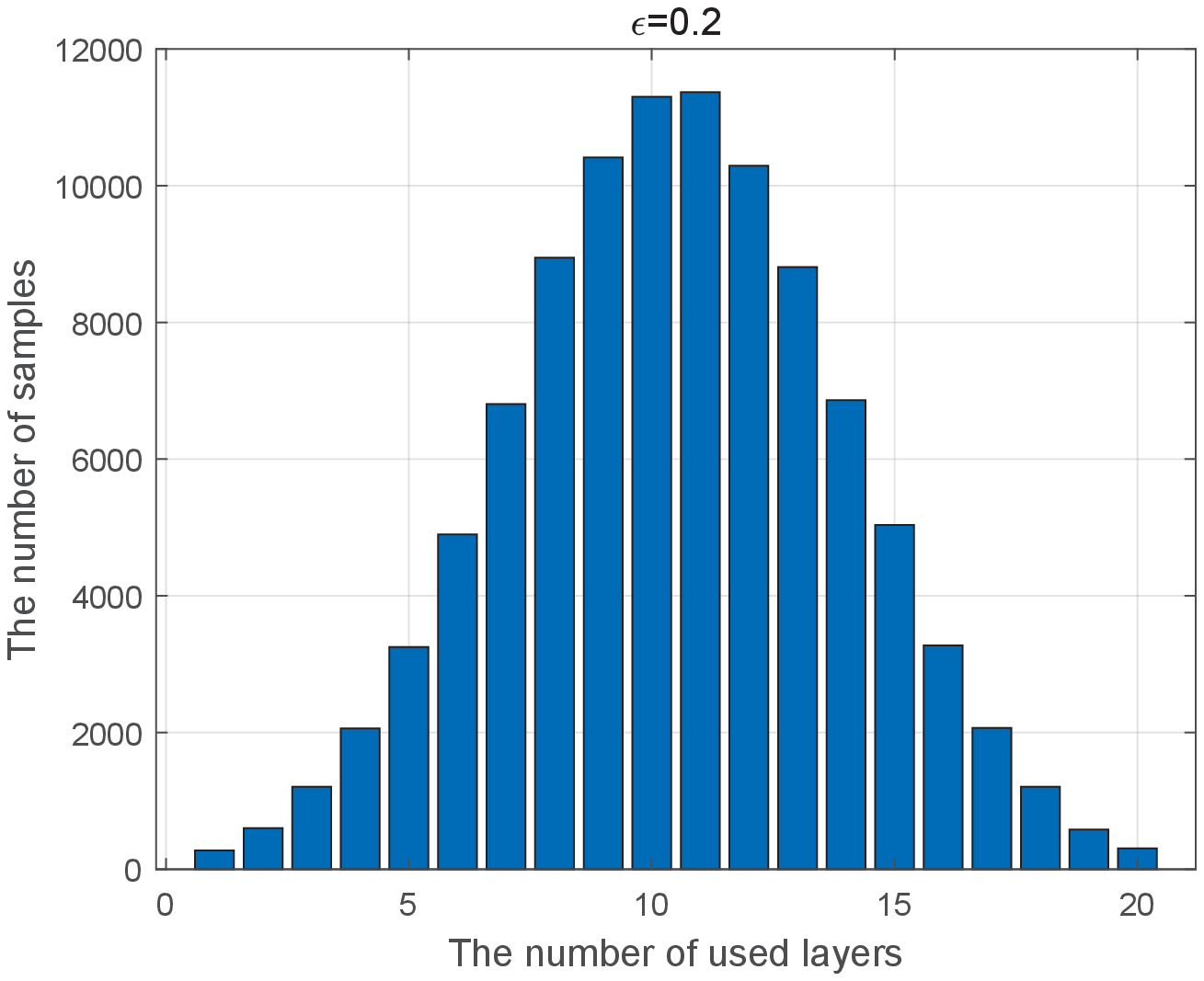}} }
\subfloat[]{\centering \scalebox{0.5}{\includegraphics{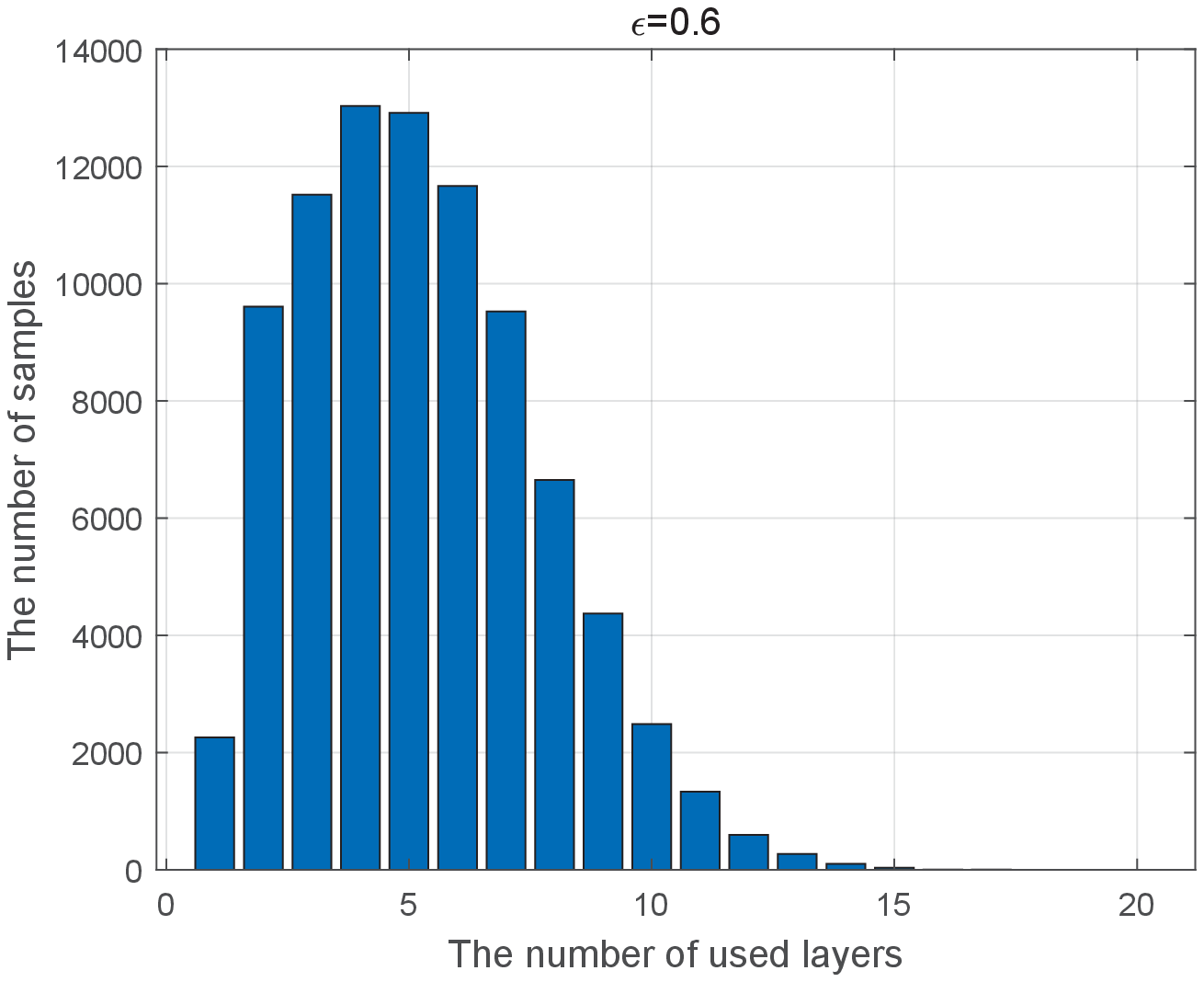}}}	
\caption{Distributions of the used number of layers with different halting constant $\epsilon$: (a) $\epsilon=0.2$; (b) $\epsilon=0.6$.}
\label{Distribution}
\end{figure*}

Fig. \ref{Distribution} shows the distributions of the used number of layers with different halting constant $\epsilon$ defined in \eqref{StopCondition}. It verifies that the adaptive layer is required for the samples with different sparsity levels. The results show that more layers are required for a smaller $\epsilon$. In particular, a smaller $\epsilon$ leads to higher precision of channel estimation but with more layers.

\begin{figure}[t]
\begin{centering}
\includegraphics[width=0.39\textwidth]{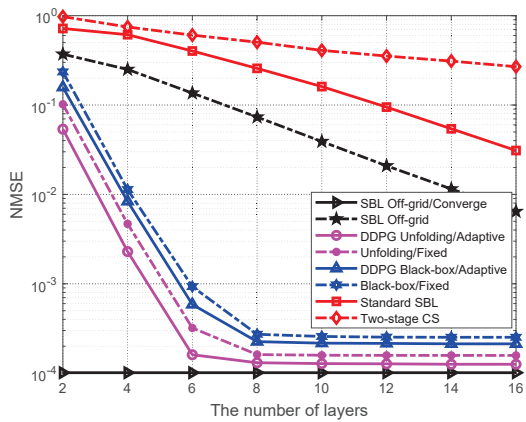}
\par\end{centering}
\caption{NMSE performance versus the number of layers $L$.}
\label{Layer}
\end{figure}

Fig. \ref{Layer} presents the NMSE performance versus the number of layers (iterations), where $L$ is the average number of layers of the whole testing dataset for the DDPG Unfolding/Adaptive and DDPG Black-box/Adaptive. The scheme ``SBL Off-grid/Converge" is considered as the lower bound since it iterates until the algorithm converges. It can be seen that the NMSE performance of all schemes decreases as the number of layers increases. For deep learning algorithms, i.e., the DDPG Unfolding/Adaptive, Unfolding/Fixed, DDPG Black-box/Adaptive, and Black-box/Fixed, the NMSE performance decreases sharply when $L$ is small and will not decrease when $L=10$, which is the optimal number of layers. In comparison, the NMSE performance achieved by the iterative optimization algorithms, i.e., the Standard SBL, Two-stage CS, and SBL Off-grid, decreases slowly with $L$, since these algorithms require a large number of layers to converge. Thus, the deep learning algorithms significantly outperform those iterative optimization algorithms when $L$ is small. Furthermore, the proposed DDPG Unfolding/Adaptive outperforms the other deep learning algorithms and achieves approaching NMSE performance to the lower bound SBL Off-grid/Converge with much reduced number of layers. Note that the optimal numbers of layers of the DDPG Unfolding/Adaptive and Unfolding/Fixed are $L=6$ and $L=8$, respectively. In other words, the proposed DDPG Unfolding/Adaptive achieves nearly the same performance as the Unfolding/Fixed with $25\%$ fewer number of layers. Thus, we can conclude that the proposed DDPG Unfolding/Adaptive achieves a satisfactory trade-off between the performance and computational complexity. 

\begin{figure}[t]
\begin{centering}
\includegraphics[width=0.39\textwidth]{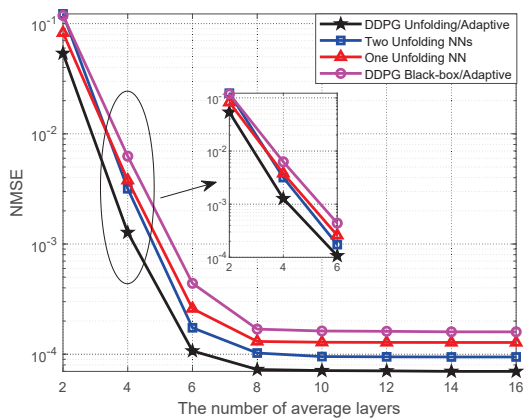}
\par\end{centering}
\caption{NMSE performance versus the number of average layers: Comparison of different deep-unfolding NN schemes.}
\label{TwoUnfold}
\end{figure}

\begin{figure*}[!t]
\centering
\subfloat[]{\centering \scalebox{0.5}{\includegraphics{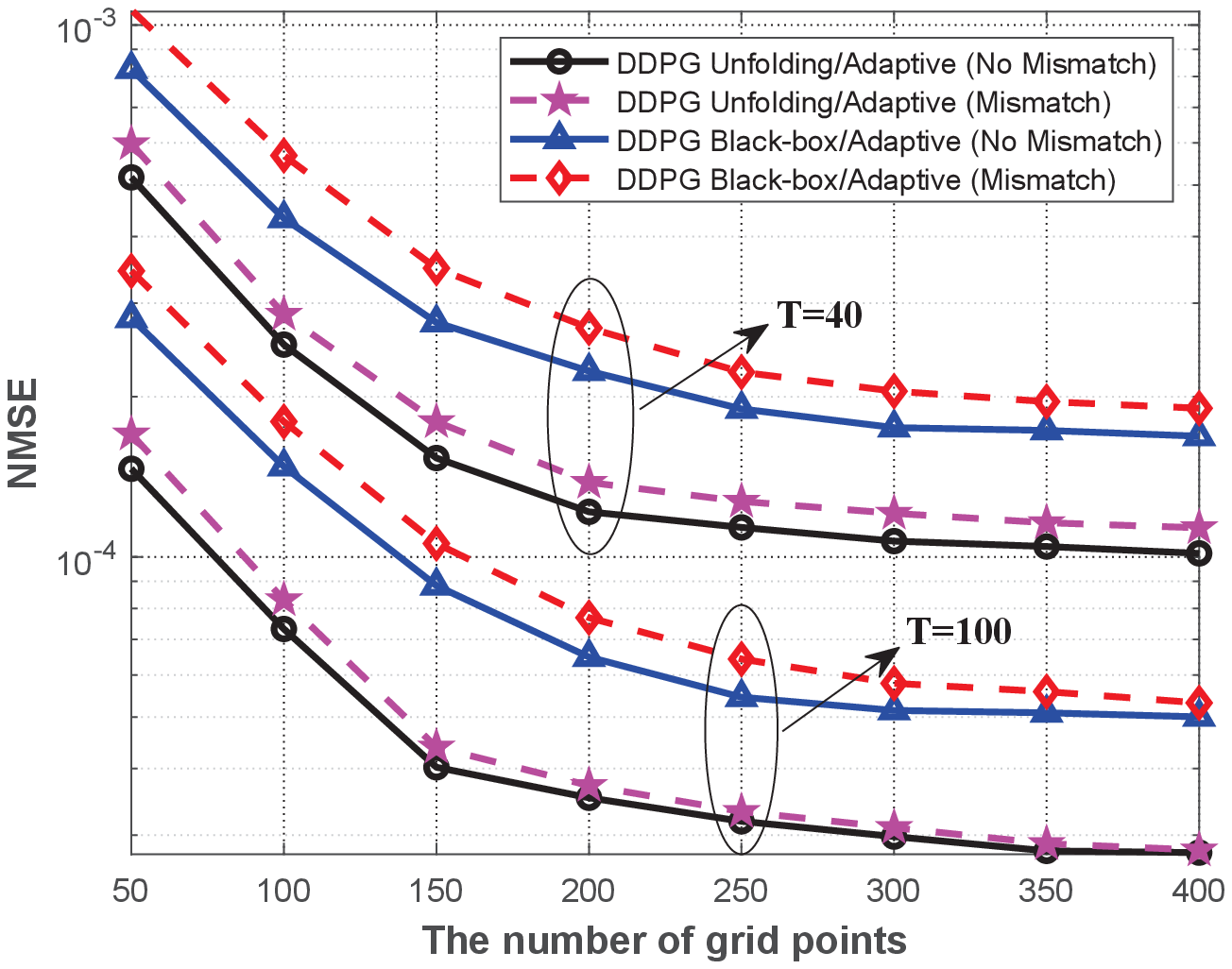}} }
\subfloat[]{\centering \scalebox{0.5}{\includegraphics{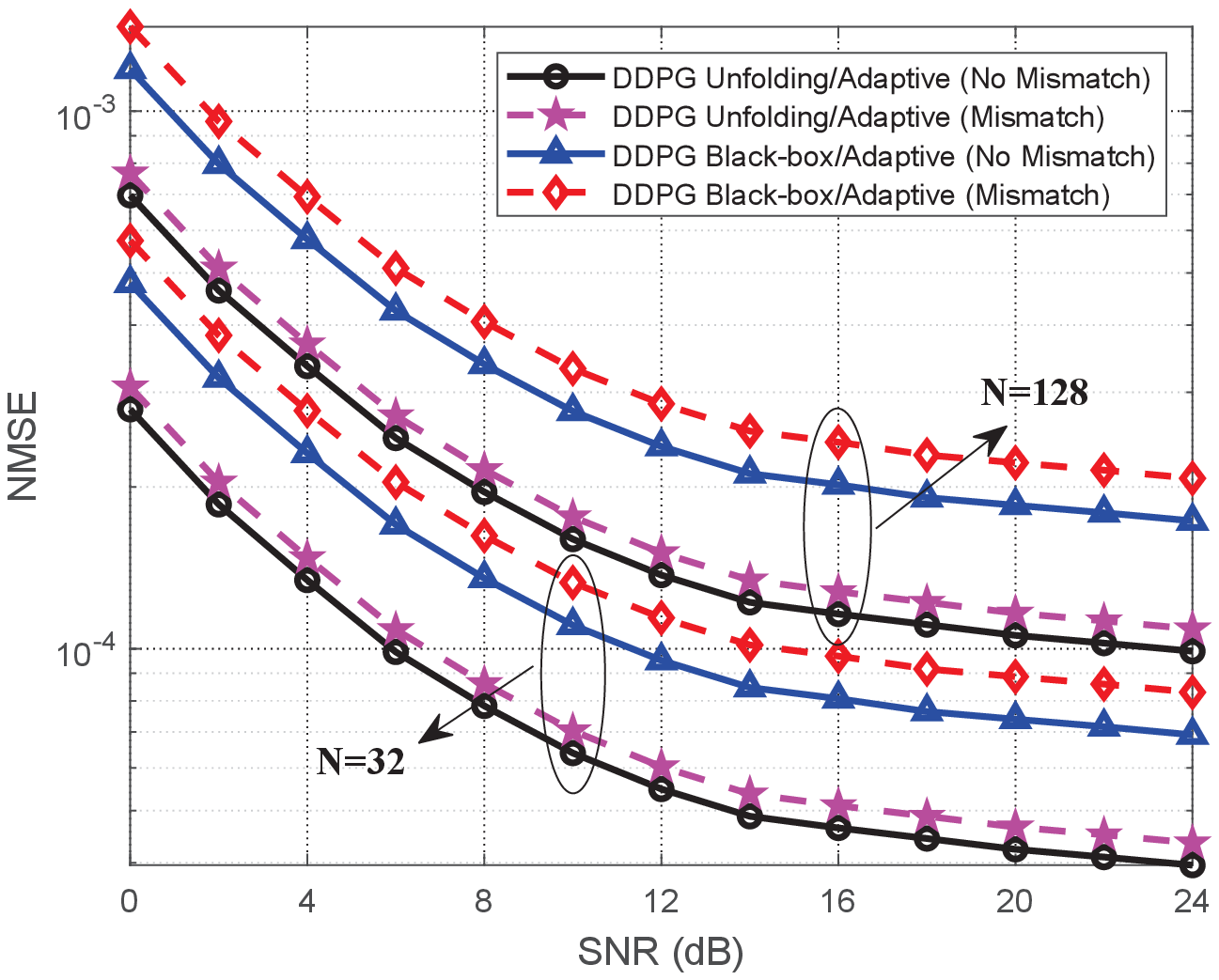}}}	
\caption{Generalization ability: (a) The number of grid points $\hat{J}$ and pilot length $T$; (b) The number of antennas $N$ and SNR. }
\label{General}
\end{figure*}

Fig. \ref{TwoUnfold} indicates the NMSE performance versus the number of average layers of the whole dataset, where four schemes are analyzed: (i) A deep-unfolding NN with fixed number of layers $L$; (ii) Two deep-unfolding NNs with fixed number of layers; (iii) DDPG-driven deep-unfolding with adaptive layers; (iv) DDPG-driven black-box NN with adaptive layers. Generally, the different number of rays $J$ leads to various sparsity levels of channel, and the samples with higher sparsity levels require much more number of layers to achieve satisfactory NMSE performance. Thus, the second scheme employs a deep-unfolding NN with fixed $L+1$ layers to deal with the samples with $13\leq J\leq 20$ and a deep-unfolding NN with fixed $L-1$ layers to handle the samples with $6\leq J<13$. It can be seen that the Two Unfolding NNs has worse NMSE performance than the One Unfolding NN when the number of layers is small. It outperforms the One Unfolding NN when $L>4$. 
Moreover, Two Unfolding NNs with $L=6$ achieves approaching performance to One Unfolding NN with $L=8$. It shows the effectiveness of handling different samples with varying number of layers. Furthermore, the NMSE performance achieved by all schemes decreases with $L$ and the proposed DDPG Unfolding/Adaptive significantly outperforms the other schemes.

\subsection{Generalization Ability}

Fig. \ref{General}(a) shows the generalization ability for the number of grid points $\hat{J}$ and pilot length $T$. We train the DDPG Unfolding/Adaptive and DDPG Black-box/Adaptive in the configuration of $\hat{J} = 400$, $T = 100$, $\textrm{SNR} = 20$ dB, and $N = 128$. The schemes in the figure with ``no mismatch" denote that the training and testing dataset have the same parameters. As for the schemes with ``mismatch", we employ the trained DDPG to test the dataset with smaller $\hat{J}$ and $T$. We can train a large DDPG and generalize it to a smaller system via zero padding. In particular, we train the DDPG with $\mathbf{X}_{1}\in \mathbb{C}^{T_{1} \times N}$ and $\mathbf{A}_{1}(\bm{\beta})=[\mathbf{a}_{1}(\hat{\phi}_{1}+\beta_{1}), \mathbf{a}_{1}(\hat{\phi}_{2}+\beta_{2}), \cdots, \mathbf{a}_{1}(\hat{\phi}_{\hat{J}_{1}}+\beta_{\hat{J}_{1}}) ]\in \mathbb{C}^{N \times \hat{J}_{1} }$, and employ it to test the dataset with smaller $\hat{J}_{2}$ and $T_{2}$. We conduct zero padding for   
$\mathbf{X}_{2}\in \mathbb{C}^{T_{2} \times N}$ and $\mathbf{A}_{2}(\bm{\beta})\in \mathbb{C}^{N \times \hat{J}_{2} }$ as $\left[\mathbf{X}_{2}; \bm{0}^{ (T_{1}-T_{2}) \times N} \right]$ and $\left[\mathbf{a}_{2}(\hat{\phi}_{1}+\beta_{1}), \mathbf{a}_{2}(\hat{\phi}_{2}+\beta_{2}), \cdots, \mathbf{a}_{2}(\hat{\phi}_{\hat{J}_{2}}+\beta_{\hat{J}_{2}}), \bm{0}^{N \times (\hat{J}_{1}-\hat{J}_{2})} \right]$, respectively, which have the same dimensions as $\mathbf{X}_{1}$ and $\mathbf{A}_{1}(\bm{\beta})$.
From the figure, it is readily seen that the performance loss for the DDPG-driven deep-unfolding is small, caused by the mismatch of $\hat{J}$ and $T$ in the training and testing stages. This demonstrates the satisfactory generalization ability of the proposed DDPG-driven deep-unfolding for different values of $\hat{J}$ and $T$. 
Moreover, the performance loss of the DDPG with mismatch decreases with $\hat{J}$ and $T$. It is because the performance loss is less when the mismatch between the training and testing configurations is smaller.

Fig. \ref{General}(b) shows the generalization ability for the number of antennas $N$ and SNR. To enhance the generalization ability for SNR, we train the DDPG Unfolding/Adaptive and DDPG Black-box/Adaptive in the configuration of $\hat{J} = 400$, $T = 100$, $\textrm{SNR} = 0, 2, \cdots, 24$ dB, and $N = 128$.  
It can be seen that the performance loss for the DDPG-driven deep-unfolding is small, caused by the mismatch of $N$ and SNR in the training and testing stages. Furthermore, the performance loss of DDPG Unfolding is smaller than that of DDPG Black-box, which demonstrates the better generalization ability of the proposed DDPG-driven deep-unfolding. 

\section{Conclusion} \label{Conclusion}
In this work, we proposed a framework of DDPG-driven deep-unfolding with adaptive depth for different inputs, where the trainable parameters are learned by the DDPG. This framework can be employed for channel estimation in massive MIMO systems. In particular, we firstly formulated the channel estimation problem with an off-grid basis and developed a SBL-based algorithm to solve it. Subsequently, the SBL-based algorithm has been unfolded into a layer-wise structure with a set of introduced trainable parameters and the performance analysis has been provided. Then, the proposed DDPG-driven deep-unfolding framework has been employed to solve this channel estimation problem based on the unfolded structure of the SBL-based algorithm. To realize the adaptive depth, we designed the halting score to indicate when to stop, which is a function of the channel reconstruction error.
Furthermore, the proposed framework has been extended to realize the adaptive depth of the general DNN. Simulation results showed that the proposed algorithm outperforms the conventional algorithms in terms of the NMSE performance with much reduced number of layers.

\begin{appendices}
\section{Proof for Theorem 1}
\label{AppendixA}

In this appendix, we prove Theorem \ref{theorem1}. We take the update of $\alpha$ in \eqref{UpdateAlpha} as an example, where $\boldsymbol{\gamma}$ and $\boldsymbol{\beta}$ are treated as constants. The update of $\boldsymbol{\gamma}$ and $\boldsymbol{\beta}$ can be analyzed similarly. Recall that
\begin{equation} \label{eta1}
\!\! \eta(\alpha, \boldsymbol{\gamma}, \boldsymbol{\beta}) \!=\! \textrm{Tr} \left( \! \boldsymbol{\Phi}(\boldsymbol{\beta}) \boldsymbol{\Sigma}(\alpha, \boldsymbol{\gamma}, \boldsymbol{\beta}) \boldsymbol{\Phi}^{H}(\boldsymbol{\beta}) \! \right) \!+\! \|\mathbf{y} \!-\! \boldsymbol{\Phi}(\boldsymbol{\beta}) \boldsymbol{\mu}(\alpha, \boldsymbol{\gamma}, \boldsymbol{\beta}) \! \|^{2},
\end{equation}
where
\begin{equation} \label{musigma1}
\begin{aligned}
\boldsymbol{\mu}(\alpha, \boldsymbol{\gamma}, \boldsymbol{\beta}) &= \alpha \boldsymbol{\Sigma}(\alpha, \boldsymbol{\gamma}, \boldsymbol{\beta}) \mathbf{\Phi}^{H}(\boldsymbol{\beta})\mathbf{y}, \\
\boldsymbol{\Sigma}(\alpha, \boldsymbol{\gamma}, \boldsymbol{\beta}) &= ( \alpha \mathbf{\Phi}^{H}(\boldsymbol{\beta}) \mathbf{\Phi}(\boldsymbol{\beta}) + \textrm{diag}(\boldsymbol{\gamma}) )^{-1}.
\end{aligned}
\end{equation}
For clarity, we denote $\boldsymbol{\Phi}(\boldsymbol{\beta})$ as $\boldsymbol{\Phi}$, and $\eta(\alpha, \boldsymbol{\gamma}, \boldsymbol{\beta})$ in \eqref{eta1} as $\eta(\alpha)$. 

Recall the closed-form solution of $\alpha$:
\begin{equation}  \label{UpdateAlpha1}
\alpha^{t+1}=\frac{T+a}{b+\eta\left(\alpha^{t} \right)}.
\end{equation}
We substitute \eqref{eta1} and \eqref{musigma1} into \eqref{UpdateAlpha1} and obtain the following mapping from $\alpha^{t}$ to $\alpha^{t+1}$:
\begin{equation}  \label{UpdateAlpha2}
\begin{aligned}
\alpha^{t+1} &= (T+a) \bigg( b+\textrm{Tr}  \left( \boldsymbol{\Phi} ( \alpha^{t} \mathbf{\Phi}^{H} \mathbf{\Phi} + \textrm{diag}(\boldsymbol{\gamma}) )^{-1} \boldsymbol{\Phi}^{H} \right) \\
& + \|\mathbf{y} - \alpha^{t} \boldsymbol{\Phi} ( \alpha^{t} \mathbf{\Phi}^{H} \mathbf{\Phi} + \textrm{diag}(\boldsymbol{\gamma}) )^{-1} \mathbf{\Phi}^{H} \mathbf{y} \|^{2} \bigg)^{-1},
\end{aligned}
\end{equation}
where $t$ denotes the index of iterations. According to \eqref{UpdateAlpha2} and the property of matrix trace: $\textrm{Tr}(\mathbf{A}\mathbf{B})=\textrm{Tr}(\mathbf{B}\mathbf{A})$, we have the following mapping from $\alpha^{t}$ to $\alpha^{t+2}$:
\begin{equation} \label{TwoIter}
\begin{aligned}  
& \alpha^{t+2} \!=\! ( \! T+a \!)  \bigg( \! b \!+\! \textrm{Tr} \big( \! \boldsymbol{\Phi}^{H} \boldsymbol{\Phi} \big( \frac{T+a}{b+\eta (\alpha^{t})} \mathbf{\Phi}^{H} \mathbf{\Phi} + \textrm{diag}(\boldsymbol{\gamma}) \big)^{-1} \! \big) \\
&  + \! \big\|  \mathbf{y} \!-\! \frac{T \!+\! a}{b \!+\! \eta (\alpha^{t})} \boldsymbol{\Phi} \big( \frac{T \!+\! a}{b \!+\! \eta (\alpha^{t})} \mathbf{\Phi}^{H} \mathbf{\Phi} \!+\! \textrm{diag}(\boldsymbol{\gamma}) \big)^{-1} \mathbf{\Phi}^{H} \mathbf{y}  \big\|^{2} \! \bigg)^{-1} \!\!\!\!\! .
\end{aligned}
\end{equation}
Recall the $l$-th layer of deep-unfolding w.r.t. $\alpha$:
\begin{equation} \label{ArchiUnfold1}
\begin{aligned}
\alpha^{l+1} &= (T+a) \bigg( b + \textrm{Tr}\left(\boldsymbol{\Phi} (\boldsymbol{\Sigma}(\alpha, \boldsymbol{\gamma}, \boldsymbol{\beta}) + \mathbf{O}_{1}) \boldsymbol{\Phi}^{H} \right)  \\
& \quad \quad \quad +\|\mathbf{y}- \boldsymbol{\Phi} (\boldsymbol{\mu}(\alpha, \boldsymbol{\gamma}, \boldsymbol{\beta}) + \mathbf{o}_{2}) \|_{2}^{2} \bigg)^{-1}.  
\end{aligned}	
\end{equation}	
By substituting \eqref{musigma1} into \eqref{ArchiUnfold1}, we obtain the mapping from $\alpha^{l}$ to $\alpha^{l+1}$: 
\begin{equation}  \label{OneLayer}
\begin{aligned} 
& \alpha^{l+1} \!=\! (T \!+\! a)\bigg( \! b \!+\! \textrm{Tr}\big( \! \boldsymbol{\Phi}^{H} \boldsymbol{\Phi} \big( \! \alpha^{l} \mathbf{\Phi}^{H} \mathbf{\Phi} \!+\! \textrm{diag}( \boldsymbol{\gamma} ) \! \big)^{-1} \!\!+\! \boldsymbol{\Phi}^{H} \boldsymbol{\Phi} \mathbf{O}_{1} \! \big)  \\
& \!+\! \big\| \mathbf{y}  \!-\! \alpha^{l} \boldsymbol{\Phi} \big( \! \big( \! \alpha^{l} \mathbf{\Phi}^{H} \mathbf{\Phi} \!+\! \textrm{diag}(\boldsymbol{\gamma}) \! \big)^{-1} \!+\! \mathbf{O}_{1} \! \big) \mathbf{\Phi}^{H} \mathbf{y} \!-\! \mathbf{\Phi}\mathbf{o}_{2} \big\|^{2} \! \bigg)^{-1} \!\!\!\!.
\end{aligned}
\end{equation}

\subsection{Deterministic Channel}
For the scenario that the channel is deterministic, e.g., changes little during the channel coherence time, it can be demonstrated that there exist trainable parameters $\mathbf{O}_{1}$ and $\mathbf{o}_{2}$ to ensure: $\alpha^{t+2}=\alpha^{l+1}$. We make the right side of \eqref{TwoIter} equal that of \eqref{OneLayer} and by solving this equation w.r.t. the variables $\mathbf{O}_{1}$ and $\mathbf{o}_{2}$, we obtain
\begin{equation}  
\begin{aligned} 
\mathbf{O}_{1} & =\! \big( \! \frac{T \!+\! a}{b \!+\! \eta (\alpha^{t})} \mathbf{\Phi}^{H} \mathbf{\Phi} \!+\! \textrm{diag}(\boldsymbol{\gamma})  \big)^{-1} \!-\! \big( \! \alpha^{l} \mathbf{\Phi}^{H} \mathbf{\Phi} \!+\! \textrm{diag}(\boldsymbol{\gamma})  \big)^{-1}, \\
\mathbf{o}_{2} &=\! \frac{T \!+\! a}{b \!+\! \eta (\alpha^{t})} \mathbf{\Phi}^{\dagger} \boldsymbol{\Phi} \big( \frac{T \!+\! a}{b \!+\! \eta (\alpha^{t})} \mathbf{\Phi}^{H} \mathbf{\Phi} \!+\! \textrm{diag}(\boldsymbol{\gamma}) \big)^{-1} \mathbf{\Phi}^{H} \mathbf{y} \\
& - \alpha^{l} \mathbf{\Phi}^{\dagger} \boldsymbol{\Phi} \big( \big( \alpha^{l} \mathbf{\Phi}^{H} \mathbf{\Phi} + \textrm{diag}(\boldsymbol{\gamma}) \big)^{-1} + \mathbf{O}_{1} \big) \mathbf{\Phi}^{H} \mathbf{y}. 
\end{aligned}
\end{equation}

\subsection{Fading Channel}
For the channel that follows a certain distribution, we need to prove the existence of $\mathbf{O}_{1}$ and $\mathbf{o}_{2}$ that makes $\mathbb{E}_{\mathbf{h}} \{ \| \alpha^{t+2}-\alpha^{l+1} \|^{2} \} < \delta$ satisfied. 
Based on \eqref{TwoIter} and \eqref{OneLayer}, we have
\begin{equation} 
\begin{aligned}  
& \mathbb{E}_{\mathbf{h}} \{ \| \alpha^{t+2}-\alpha^{l+1} \| \} \\
= & \mathbb{E}_{\mathbf{h}} \{ \| ( T+a ) A^{-1} - (T + a) B^{-1} \| \} \\
= & \mathbb{E}_{\mathbf{h}} \{ \| \frac{(T+a)}{AB}(A-B) \| \} \\
\overset{(a)}{\leq} & \big(\mathbb{E}_{\mathbf{h}} \{ \| \frac{(T+a)}{AB}\|^{2} \} \big)^{\frac{1}{2}} \big( \mathbb{E}_{\mathbf{h}} \{ \|A-B\|^{2} \} \big)^{\frac{1}{2}}, 
\end{aligned}
\end{equation}
where the inequality $(a)$ is due to \textit{The Cauchy-Schwarz Inequality} and we denote 
\begin{equation} 
\begin{aligned} 
& A \triangleq  b + \textrm{Tr} \big( \boldsymbol{\Phi}^{H} \boldsymbol{\Phi} \big( \frac{T+a}{b+\eta (\alpha^{t})} \mathbf{\Phi}^{H} \mathbf{\Phi} + \textrm{diag}(\boldsymbol{\gamma}) \big)^{-1} \big) \\
& + \big\|  \mathbf{y} - \frac{T + a}{b + \eta (\alpha^{t})} \boldsymbol{\Phi} \big( \frac{T + a}{b + \eta (\alpha^{t})} \mathbf{\Phi}^{H} \mathbf{\Phi} + \textrm{diag}(\boldsymbol{\gamma}) \big)^{-1} \mathbf{\Phi}^{H} \mathbf{y}  \big\|^{2}, \\
& B \triangleq b + \textrm{Tr}\big( \boldsymbol{\Phi}^{H} \boldsymbol{\Phi} \big(  \alpha^{l} \mathbf{\Phi}^{H} \mathbf{\Phi} + \textrm{diag}( \boldsymbol{\gamma} ) \big)^{-1} + \boldsymbol{\Phi}^{H} \boldsymbol{\Phi} \mathbf{O}_{1} \big)  \\
& + \big\| \mathbf{y} - \alpha^{l} \boldsymbol{\Phi} \big( \big( \alpha^{l} \mathbf{\Phi}^{H} \mathbf{\Phi} + \textrm{diag}(\boldsymbol{\gamma}) \big)^{-1} + \mathbf{O}_{1} \big) \mathbf{\Phi}^{H} \mathbf{y} - \mathbf{\Phi}\mathbf{o}_{2} \big\|^{2}.
\end{aligned}
\end{equation}

\begin{figure*}
\begin{subequations} \label{FirstItemab} 
\begin{eqnarray} 
& &\!\!\!\!\!\! \mathbb{E}_{\mathbf{h}}  \{ \|A-B\|^{2} \} \label{FirstItem1} = \mathbb{E}_{\mathbf{h}} \bigg\{  \bigg\| \textrm{Tr} \bigg( \mathbf{\Phi}^{H} \mathbf{\Phi} \bigg( \big( \frac{T + a}{b + \eta (\alpha^{t})} \mathbf{\Phi}^{H} \mathbf{\Phi} + \textrm{diag} ( \boldsymbol{\gamma})  \big)^{-1} - \big( \alpha^{l} \mathbf{\Phi}^{H} \mathbf{\Phi} + \textrm{diag}( \boldsymbol{\gamma}) \big)^{-1} - \mathbf{O}_{1} \bigg) \bigg) + \big\|  \mathbf{y} \notag \\
&-&\!\!\!\!\!\! \frac{T + a}{b + \eta (\alpha^{t})} \boldsymbol{\Phi} \big( \frac{T + a}{b + \eta (\alpha^{t})} \mathbf{\Phi}^{H} \mathbf{\Phi} \!+\! \textrm{diag}(\boldsymbol{\gamma}) \big)^{-1} \mathbf{\Phi}^{H} \mathbf{y} \big\|^{2} \!-\! \big\| \mathbf{y} \!-\! \alpha^{l} \boldsymbol{\Phi} \big( \big( \alpha^{l} \mathbf{\Phi}^{H} \mathbf{\Phi} \!+\! \textrm{diag}(\boldsymbol{\gamma}) \big)^{-1} \!+\! \mathbf{O}_{1} \big) \mathbf{\Phi}^{H} \mathbf{y} \!-\! \mathbf{\Phi}\mathbf{o}_{2} \big\|^{2} \bigg\|^{2} \! \bigg\} \label{FirstItema} \\
& \overset{(b)}{\leq} &\!\!\!\!\!\! \mathbb{E}_{\mathbf{h}} \bigg\{  \bigg\| \textrm{Tr} \bigg(  \big( \frac{T + a}{b + \eta (\alpha^{t})} \mathbf{\Phi}^{H} \mathbf{\Phi} + \textrm{diag} ( \boldsymbol{\gamma})  \big)^{-1} - \big( \alpha^{l} \mathbf{\Phi}^{H} \mathbf{\Phi} + \textrm{diag}( \boldsymbol{\gamma}) \big)^{-1} - \mathbf{O}_{1} \bigg) \bigg\|^{2} \bigg\} \notag \\
&+&\!\!\!\!\!\! \mathbb{E}_{\mathbf{h}} \bigg\{  \bigg\| \frac{T + a}{b + \eta (\alpha^{t})} \boldsymbol{\Phi} \big( \frac{T + a}{b + \eta (\alpha^{t})} \mathbf{\Phi}^{H} \mathbf{\Phi} \!+\! \textrm{diag}(\boldsymbol{\gamma}) \big)^{-1} \mathbf{\Phi}^{H} \mathbf{y} - \alpha^{l} \boldsymbol{\Phi} \big( \big( \alpha^{l} \mathbf{\Phi}^{H} \mathbf{\Phi} + \textrm{diag}(\boldsymbol{\gamma}) \big)^{-1} + \mathbf{O}_{1} \big) \mathbf{\Phi}^{H} \mathbf{y} - \mathbf{\Phi}\mathbf{o}_{2} \bigg\|^{2} \! \bigg\} \label{FirstItem}
\end{eqnarray}
\end{subequations}
\end{figure*}

We focus on the term $\mathbb{E}_{\mathbf{h}} \{ \|A-B\|^{2} \}$ and obtain \eqref{FirstItemab}. Note that the inequality $(b)$ is due to \textit{The Cauchy-Schwarz Inequality}, \textit{The Triangle Inequality}, and \textit{The Absolute Value Inequality}. Then, we set
\begin{equation} \label{O1}
\mathbf{O}_{1} \!\!=\! \mathbb{E}_{\mathbf{h}} \! \bigg\{ \! \big( \! \frac{T \!+\! a}{b \!+\! \eta (\alpha^{t})} \mathbf{\Phi}^{H} \mathbf{\Phi} + \textrm{diag}(\boldsymbol{\gamma}) \! \big)^{-1} \!-\! \big( \alpha^{l} \mathbf{\Phi}^{H} \mathbf{\Phi} \!+\! \textrm{diag}(\boldsymbol{\gamma}) \! \big)^{-1} \! \bigg\}. 
\end{equation}
Then, we propose the following Lemma \ref{lemma1} as:
\begin{lemma} \label{lemma1}
According to \textit{The Law of Large Numbers}, the first item of \eqref{FirstItem} converges to $0$ with a sufficiently large number of channel samples. 
\end{lemma}

The proof of Lemma \ref{lemma1} is provided in Appendix \ref{AppendixB}. Thus, $\delta$ only depends on the second item of \eqref{FirstItem} as 
\begin{equation}  \label{SecondItem} 
\begin{aligned} 
& \mathbb{E}_{\mathbf{h}} \bigg\{ \bigg\| 
\frac{T+a}{b+\eta (\alpha^{t})} \boldsymbol{\Phi} \big( \frac{T+a}{b+\eta (\alpha^{t})} \mathbf{\Phi}^{H} \mathbf{\Phi} + \textrm{diag}(\boldsymbol{\gamma}) \big)^{-1} \mathbf{\Phi}^{H} \mathbf{y} \\
& -  \alpha^{l} \boldsymbol{\Phi} \big( \big( \alpha^{l} \mathbf{\Phi}^{H} \mathbf{\Phi} + \textrm{diag} (\boldsymbol{\gamma}) \big)^{-1} + \mathbf{O}_{1} \big) \mathbf{\Phi}^{H} \mathbf{y} -\mathbf{\Phi}\mathbf{o}_{2}
\bigg\|^{2} \bigg\} \\
\overset{(c)}{\leq} & \mathbb{E}_{\mathbf{h}} \bigg\{ \bigg\| \frac{T+a}{b+\eta (\alpha^{t})} \big( \frac{T+a}{b+\eta (\alpha^{t})} \mathbf{\Phi}^{H} \mathbf{\Phi} + \textrm{diag}(\boldsymbol{\gamma}) \big)^{-1} \mathbf{\Phi}^{H} \mathbf{y} \\
& -  \alpha^{l} \mathbb{E}_{\mathbf{h}}   \big\{ \big( \frac{T+a}{b+\eta (\alpha^{t})} \mathbf{\Phi}^{H} \mathbf{\Phi} + \textrm{diag}(\boldsymbol{\gamma}) \big)^{-1} \big\} \mathbf{\Phi}^{H} \mathbf{y} - \mathbf{o}_{2}
\bigg\|^{2} \bigg\},
\end{aligned}
\end{equation}
where the inequality $(c)$ is obtained by substituting the expression of $\mathbf{O}_{1}$ in \eqref{O1} and based on \textit{The Absolute Value Inequality}. Then, we set
\begin{equation} 
\begin{aligned}
& \mathbf{o}_{2}=\mathbb{E}_{\mathbf{h}} \big\{ \big\| 
\frac{T+a}{b+\eta (\alpha^{t})} \big( \frac{T+a}{b+\eta (\alpha^{t})} \mathbf{\Phi}^{H} \mathbf{\Phi} + \textrm{diag}(\boldsymbol{\gamma}) \big)^{-1} \mathbf{\Phi}^{H} \mathbf{y} \\
& \quad - \alpha^{l} \mathbb{E}_{\mathbf{h}}   \big\{ \big( \frac{T+a}{b+\eta (\alpha^{t})} \mathbf{\Phi}^{H} \mathbf{\Phi} + \textrm{diag}(\boldsymbol{\gamma}) \big)^{-1} \big\} \mathbf{\Phi}^{H} \mathbf{y} \big\|^{2} \big\}. 
\end{aligned}
\end{equation}
According to \textit{The Law of Large Numbers}, \eqref{SecondItem} converges to $0$ with a sufficiently large number of channel samples, which can be proved in a similar way to Lemma \ref{lemma1}. Thus, we can conclude that there exist trainable parameters to ensure the difference between $\alpha^{t+2}$ and $\alpha^{l+1}$ to be smaller than a sufficiently small value $\delta$, i.e., $\mathbb{E}_{\mathbf{h}} \{ \| \alpha^{t+2}-\alpha^{l+1} \| \} < \delta$.

\section{ Proof for Lemma \ref{lemma1} }
\label{AppendixB}
In this appendix, we prove Lemma \ref{lemma1}. Recall \textit{The Law of Large Numbers} as:

\begin{theorem}
Denote $X$ as a random variable that follows a certain distribution and $x_{n}, n=1,2, \cdots, N$ as $N$ samples of $X$. The average of a large number of samples, $\frac{1}{N}\sum_{n=1}^{N} x_{n}$, approaches to the expectation of $X$, $\mathbb{E}(X)$, and tends to become closer to $\mathbb{E}(X)$ with the increase of $N$, which can be written as
\begin{equation} \label{LLN0}
\lim\limits_{N\rightarrow \infty} P\bigg( \bigg\| \frac{1}{N}\sum\limits_{n=1}^{N} x_{n} - \mathbb{E}(X) \bigg\|<\varepsilon \bigg)=1,
\end{equation}
where $\varepsilon>0$ is a small enough number. 
\end{theorem}

Recall the first item of \eqref{FirstItem} as
\begin{equation} \label{Firstitem2}
\begin{aligned} 
& \mathbb{E}_{\mathbf{h}} \bigg\{  \bigg\| \big( \frac{T + a}{b + \eta (\alpha^{t})} \mathbf{\Phi}^{H} \mathbf{\Phi} + \textrm{diag} ( \boldsymbol{\gamma})  \big)^{-1} \\
& \quad \quad \quad \quad \quad - \big(  \alpha^{l} \mathbf{\Phi}^{H} \mathbf{\Phi} + \textrm{diag}( \boldsymbol{\gamma})  \big)^{-1} - \mathbf{O}_{1} \bigg\|^{2} \bigg\}.
\end{aligned} 
\end{equation}
We denote
\begin{equation}
F(\mathbf{h})\triangleq \! \big( \! \frac{T + a}{b + \eta (\alpha^{t})} \mathbf{\Phi}^{H} \mathbf{\Phi} + \textrm{diag} ( \boldsymbol{\gamma})  \big)^{-1} \!-\! \big( \alpha^{l} \mathbf{\Phi}^{H} \mathbf{\Phi} + \textrm{diag}( \boldsymbol{\gamma}) \big)^{-1}
\end{equation}
which is a function of the input sample, $\mathbf{h}$. Then, we set the trainable parameter $\mathbf{O}_{1}$ as $\mathbf{O}_{1}=\mathbb{E}_{\mathbf{h}} \{F(\mathbf{h})\}$. Then, \eqref{Firstitem2} can be rewritten as
\begin{equation} \label{LLN1}
\mathbb{E}_{\mathbf{h}} \bigg\{ \bigg\| F(\mathbf{h}) - \mathbb{E}_{\mathbf{h}} \{F(\mathbf{h})\} \bigg\|^{2} \bigg\}.
\end{equation}
Note that we employ a large number of training samples to train the deep-unfolding NN and its performance is evaluated by the average performance of a large number of testing samples. Thus, $F(\mathbf{h})$ is approximated by the average of a large number of samples, i.e., $\frac{1}{N} \sum_{n=1}^{N} F(\mathbf{h}_{n})$, in the proposed deep-unfolding NN, where $N$ denotes the number of samples that is sufficiently large. Then, we can rewrite \eqref{LLN1} as
\begin{equation} 
\mathbb{E}_{\mathbf{h}} \bigg\{ \bigg\| \frac{1}{N} \sum\limits_{n=1}^{N} F(\mathbf{h}_{n}) - \mathbb{E}_{\mathbf{h}} \{F(\mathbf{h})\} \bigg\|^{2} \bigg\}.
\end{equation}
Based on \textit{The Law of Large Numbers} in \eqref{LLN0}, we have
\begin{equation} 
\lim\limits_{N\rightarrow \infty} P\bigg( \bigg\| \frac{1}{N} \sum\limits_{n=1}^{N} F(\mathbf{h}_{n}) - \mathbb{E}_{\mathbf{h}} \{F(\mathbf{h})\} \bigg\|^{2}<\varepsilon^{2} \bigg)=1.
\end{equation}
Thus, we can conclude that \eqref{LLN1} converges to $0$ with probability $1$.   

\end{appendices}

\bibliographystyle{IEEEtran}
\bibliography{IEEEabrv,DRL4Unfold}

\end{document}